\pdfoutput=1
\documentclass[journal]{IEEEtran}

\usepackage{cite}
\usepackage{algorithm,algorithmic}
\usepackage{ulem}
\usepackage{booktabs}
\usepackage{amsmath}
\usepackage{newtxmath}
\usepackage{array}
\usepackage{graphicx}
\usepackage{subcaption}
\usepackage{url}
\usepackage{hyperref}
\usepackage{color}
\usepackage{stackengine}
\usepackage{mathrsfs}
\DeclareMathAlphabet{\mathcal}{OMS}{cmsy}{m}{n}
\usepackage{newtxtext}
\usepackage{xcolor}
\DeclareSymbolFont{matha}{OML}{txmi}{m}{it}% txfonts

\usepackage{booktabs}
\usepackage{caption}
\usepackage{fancyhdr}
\usepackage{lipsum}
\fancypagestyle{plain}{\pagestyle{fancy}}

\begin{document}

\pagestyle{fancy}
\fancyhead[C]{\fontsize{6}{8} \fontfamily{phv}\selectfont This article has been accepted for publication in IEEE Transactions on Power Systems. This is the author's version which has not been fully edited and content may change prior to final publication. Citation information: DOI 10.1109/TPWRS.2021.3120195}
\fancyfoot[C]{\fontsize{6}{6} \fontfamily{phv}\selectfont © 2021 IEEE. Personal use is permitted, but republication/redistribution requires IEEE permission. See https://www.ieee.org/publications/rights/index.html for more information.}
\fancyhead[R]{\thepage}
\title{Adaptive Coalition Formation-Based Coordinated Voltage Regulation in Distribution Networks}
\author{Yao Long,~\IEEEmembership{Student Member,~IEEE,}
        Ryan T. Elliott,~\IEEEmembership{Member,~IEEE,}
        Daniel S. Kirschen,~\IEEEmembership{Fellow,~IEEE}% <-this % stops a space
\thanks{Y. Long, R. T. Elliott, and D. S. Kirschen are with the Department of Electrical and Computer Engineering, University of Washington, Seattle, WA, 98195 USA (e-mail: longyao@uw.edu; ryanelliott@ieee.org; kirschen@uw.edu). \textit{(Corresponding author: Yao Long)}}}
\newcommand\ubar[1]{\stackunder[1.2pt]{$#1$}{\rule{.8ex}{.075ex}}}
%\markboth{}%

\maketitle \thispagestyle{fancy} 
\begin{abstract}
High penetrations of photovoltaic (PV) systems can cause severe voltage quality problems in distribution networks. This paper proposes a distributed control strategy based on the dynamic formation of coalitions to coordinate a large number of PV inverters for voltage regulation. In this strategy, a rule-based coalition formation scheme deals with the zonal voltage difference caused by the uneven integration of PV capacity. Under this scheme, PV inverters form into separate voltage regulation coalitions autonomously according to local, neighbor as well as coalition voltage magnitude and regulation capacity information. To coordinate control within each coalition, we develop a feedback-based leader-follower consensus algorithm which eliminates the voltage violations caused by the fast fluctuations of load and PV generation. This algorithm allocates the required reactive power contribution among the PV inverters according to their maximum available capacity to promote an effective and fair use of the overall voltage regulation capacity. Case studies based on realistic distribution networks and field-recorded data validate the effectiveness of the proposed control strategy. Moreover, comparison with a centralized network decomposition-based scheme shows the flexibility of coalition formation in organizing the distributed PV inverters. The robustness and generalizability of the proposed strategy are also demonstrated.
\end{abstract}

\begin{IEEEkeywords}
% Note that keywords are not normally used for peer review papers.
Consensus algorithm, distributed control, photovoltaic generation, smart inverter, voltage regulation.
\end{IEEEkeywords}

\IEEEpeerreviewmaketitle

\section{Introduction}
\IEEEPARstart{I}{nstallations} of rooftop photovoltaic (PV) systems are growing rapidly. In 2019, around 42 GW of rooftop PV systems were installed worldwide and SolarPower Europe estimated that the global installed capacity will rise to 65 GW by 2024 \cite{SolarPowerEurope}. Because they have a much larger R/X ratio than transmission networks, voltages in distribution networks are quite sensitive to the active power generated by distributed PV systems. Distribution networks with a high penetration of PV are therefore likely to experience several voltage quality problems\cite{Voltagerise, Voltagestability, Voltagedifference}:
\begin{itemize}
    \item Over-voltages when the PV generation exceeds the local load and the power flow reverses;
    \item Rapid voltage fluctuations caused by sudden changes in solar irradiance;
    \item Large voltage differences between nodes due to the uneven distribution of PV generation capacity.
\end{itemize}

Traditional voltage regulation relies on devices such as step voltage regulators (SVR), on-load tap-changing (OLTC) transformers, and capacitor banks (CB). However, these devices are intended to manage voltage variations on a time-scale of hours, and they are not able to deal with the rapid fluctuations in  the output of PV systems \cite{Traditional1, Traditional2}. On the other hand, PV inverters are capable of providing fast and flexible reactive power compensation. IEEE standard 1547 standard encourages their contribution to voltage regulation \cite{1547}.

\subsection{Literature Review}
Existing voltage regulation architectures can be broadly classified into three categories. Control schemes of the first category, such as the droop-based Volt/VAR control (VVC) \cite{Local}, rely solely on local measurements. Because they lack a global perspective and coordination, these controllers may not deploy the available resources effectively. The second category involves a centralized architecture, where a central controller with full knowledge of the network dispatches all controllable devices \cite{Centralized}. While theoretically optimal, such an approach requires extensive communications and involves complex computations. Moreover, it makes the system susceptible to a single point of failure. Distributed control architectures make up the third category. In this approach, individual controllers cooperate using communications with a limited number of neighbors to achieve a common goal. The computation and communication burden is divided among the individual controllers, which makes this approach more robust to communication failures and better able to coordinate the control of distributed voltage regulation devices in real-time.

Several authors have approached distributed voltage control as an optimization problem and implemented algorithms such as dual-ascent \cite{Dopt1}, primal-dual gradient \cite{Dopt2} to solve it in a distributed manner. While these optimization-based control strategies provide convergence and optimality guarantees under assumptions of power flow linearity and static power consumption, they usually require knowledge of the exact values of the system parameters, such as line impedances, which are not always available for distribution networks. Several model-free control strategies have also been proposed to coordinate local droop-controlled voltage regulation devices, among them strategies based on consensus algorithms. Using peer-to-peer communication, these strategies aim to allocate the required control actions, e.g. reactive power compensation or active power curtailment, fairly among PV inverters based on their available capacity  \cite{Consensus1, Consensus2, Consensus3, Consensus4, Consensus5, Consensus6}. In \cite{Consensus1, Consensus2, Consensus3}, a leader inverter measures its local voltage and updates its utilization ratio, which is the ratio of the inverter's power output used for voltage regulation to its maximum available power capacity. This ratio is delivered to all the follower inverters via neighboring communication according to the leader-follower consensus algorithm. Every follower inverter accepts this utilization ratio and adjusts its power output accordingly. After the leader inverter stops updating and the utilization ratio has been fully communicated, every participating inverter shares an equal utilization ratio. Some authors \cite{Consensus1, Consensus2} make the assumption that the highest or lowest voltage always occurs at the end node of the radial feeder and choose the inverter at this node to be the leader. However, this assumption does not always hold true when PV capacity is unevenly distributed along the feeder. Zeraati et al. \cite{Consensus3} extended this approach to let any PV inverter that can experience a voltage limit violation act as the leader. The need to appoint a leader is relaxed in \cite{Consensus4, Consensus5, Consensus6}. In those control strategies, each inverter measures the local voltage and determines its own utilization ratio. This ratio is gradually adjusted based on the consensus algorithm to reflect its neighbors' contribution to voltage regulation.

\subsection{Paper Contribution and Organization}
This paper considers the same voltage regulation objectives as in \cite{Consensus1, Consensus2, Consensus3, Consensus4}, i.e. maintaining the voltage magnitude within a satisfactory range and equalizing the utilization ratio of the participating inverters. These two objectives can conflict because equalizing the utilization ratio means that voltage violations are eliminated by increasing or decreasing the voltage across the entire network. However, if the PV generation capacity and the load are distributed unevenly, simultaneous upper and lower voltage limit violations can occur on a feeder. In this case, requiring an equal utilization ratio would either solve the over-voltage problem at certain nodes at the expense of making the under-voltage problem at other nodes even worse, or vice-versa. Therefore, a system-wide consensus on the utilization ratio becomes detrimental. To fairly share the voltage regulation burden among PV inverters while satisfying the voltage constraints, one must be able to distinguish between situations where reaching a system-wide consensus is desirable and situations where the inverters should reach a localized consensus about an equal utilization ratio. In multi-agent control, a coalition refers to a goal-directed and short-lived group formed by  smart agents to complete a common task cooperatively\cite{Paradigm}. PV inverters equipped with sensing, communicating and computing capabilities can be treated as smart agents who can work together at regulating voltages \cite{Smart1, Smart2}. These smart PV inverters can be empowered to form coalitions that can reach an effective consensus about what needs to be done to regulate voltages.

This paper proposes a distributed scheme where PV inverters organize themselves into different voltage regulation coalitions using local decision-making based on voltage magnitude as well as regulation capacity information. Within each coalition, a feedback-based leader-follower consensus algorithm determines the utilization ratio of each inverter. The main contributions of this paper are:

\begin{itemize}
    \item An adaptive coalition formation scheme which effectively solves potential conflicts between voltage regulation and reactive power sharing. In this scheme, PV inverters use a simple and clear decision logic to determine the scope of their cooperation. They can therefore organize themselves efficiently and flexibly to adapt to constantly varying network operating condition.
    \item An improved leader-follower consensus algorithm which effectively eliminates voltage violations and equalizes the utilization ratios in each coalition. Instead of relying on a fixed leader, this algorithm enables each PV inverter to adaptively determine whether it is a leader or a follower based on its voltage deviation. The connection with the distributed optimization algorithm demonstrates the theoretical foundation of this algorithm.
    \item A model-free two-timescale distributed control strategy which fully exploits the autonomous capability of the smart PV inverters to regulate voltages. This strategy enables each inverter to act in response to real-time network conditions, and does not require central control or prior knowledge of the network. Hence, it is not only able to adapt to changing network conditions and configurations, but can also scale to accommodate an arbitrary number of inverters. We demonstrate the robustness and generalizability of the proposed strategy using simulation-based case studies.
\end{itemize}

The remainder of this paper is organized as follows. Section II introduces the overall control framework. Sections III and IV describe the details of the voltage regulation coalition formation scheme and the within-coalition coordination. Section V discusses the implementation issues. Section VI and VII present and analyze simulation results. Section VIII concludes.

\section{Overview of the Proposed Control Strategy}
\subsection{Communication Network}
This paper treats the distribution network as a cyber-physical system, where each PV inverter is able to:
\begin{itemize}
    \item Acquire local measurements;
    \item Communicate with its neighbors;
    \item Compute necessary control actions;
    \item Execute these control actions. 
\end{itemize}

Communication within a network of $n$ PV inverters can be represented by a directed graph ${\mathcal{G} = (\mathcal{V}, \mathcal{E})}$, where  ${\mathcal{V} = \{1, 2,..., n\}}$ is the set of nodes and ${\mathcal{E}\subseteq \mathcal{V}\times \mathcal{V}}$ the set of edges. Each node corresponds to a PV inverter and ${(i, j) \in \mathcal{E}}$ if inverter $i$ can receive information from inverter $j$. The graph is \textit{connected} if for every pair of nodes there is a path connecting them. The graph is \textit{undirected} if ${(i, j) \in \mathcal{E}}$ implies ${(j, i) \in \mathcal{E}}$. The set of neighbors of inverter $i$ is denoted by ${\mathcal{N}_i = \{j \in \mathcal{V}: (i, j) \in \mathcal{E} \}}$, which consists of inverters that are directly connected to it through power lines without passing other inverters. Specifically, if two inverters are installed on different laterals, they are neighbors if they do not have a common upstream or downstream inverter. For example, one could assume that the inverters communicate with their neighbors using Power Line Communication (PLC) \cite{PLC}. Therefore, the communication network for a typical distribution network is a connected and undirected graph with a tree topology.

\subsection{Control Objectives}
The first control objective is to maintain the voltages within the regulation range based on the grid code:
\begin{equation}
     \ubar{V} \le V_i \le \bar{V}
 \end{equation}
where $V_i$ is the voltage magnitude of node $i$, $\ubar{V}$ and $\bar{V}$ are the lower and upper voltage limits. 

To prevent the excessive use and early saturation of the voltage regulation devices at certain nodes, several authors \cite{Consensus1, Consensus2, Consensus3, Consensus4} argue that all devices should operate at the same utilization ratio after reaching equilibrium. However, voltages in some parts of a distribution feeder may exceed the upper limit while voltages in other parts may be below the lower limit. To address this problem, we develop a scheme that allows the PV inverters to self-organize into coalitions. Within each coalition, all inverters operate at the same utilization ratio:
\begin{equation} \label{eq: utilization}
     \frac{Q_{i}}{Q_{i}^{\mathrm{max}}} = \frac{Q_{j}}{Q_{j}^{\mathrm{max}}} = u_a, \;
     \forall{i, j \in \mathcal{G}_a}
\end{equation}
where $u_a$ is the utilization ratio of coalition ${\mathcal{G}_a = (\mathcal{V}_a, \mathcal{E}_a)}$, ${\mathcal{V}_a \subseteq \mathcal{V}}$ and ${\mathcal{E}_a \subseteq \mathcal{E}}$. $Q_{i}$ and $Q_{i}^{\mathrm{max}}$ are the required reactive power contribution for voltage regulation and the maximum available reactive power capacity of PV inverter $i$.

\subsection{Control Framework}

\begin{figure}[ht!]
\centering
\includegraphics[width=3.5in]{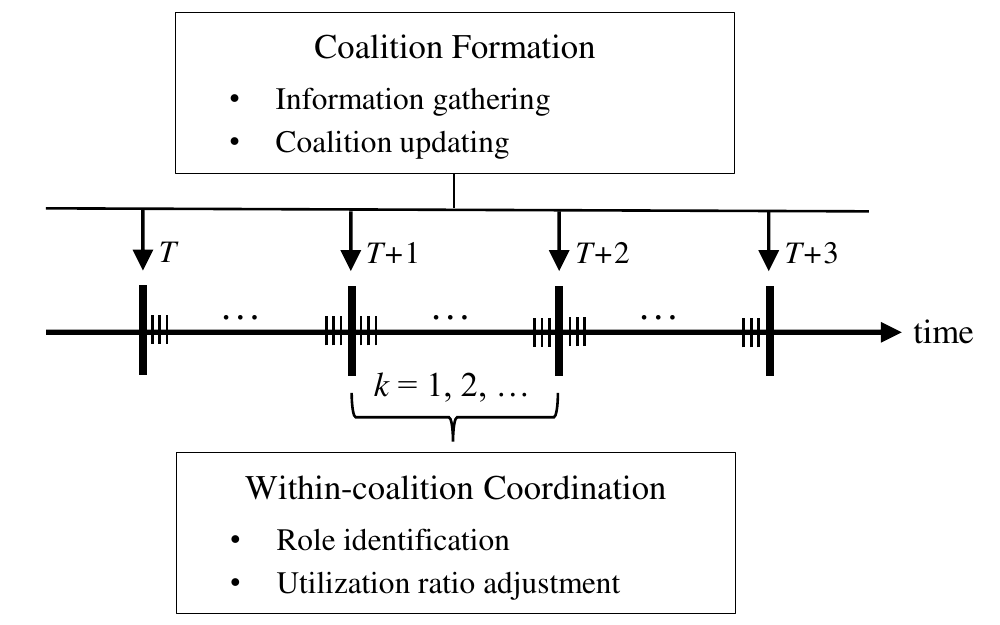}
\caption{Framework of the proposed control strategy.}
\label{FlowChart}
\end{figure}

Fig.~\ref{FlowChart} illustrates the two time-scale voltage regulation strategy. Due to the uneven distribution of load and PV generation capacity, their regular daily variations can lead to large zonal voltage differences at certain times. On the slower time-scale, e.g. every 5 minutes, the coalition formation scheme guides PV inverters to form into separate groups to solve the over-voltage and under-voltage problems that may arise simultaneously in different parts of the network. On the faster time-scale (i.e., the inverter sampling period  of tens to hundreds of milliseconds \cite{Consensus1, Consensus2, Consensus3, Consensus4, Consensus5, Consensus6}), the feedback-based leader-follower consensus algorithm eliminates the real-time voltage violations and distributes the voltage regulation burden fairly to every inverter within each coalition. Throughout the process, PV inverters make control decisions based on simple calculations and only need to exchange information with a limited number of neighbors. This control strategy is thus able to coordinate the PV inverters in real-time. Moreover, since there is no need for system information such as line impedance, network topology or nodal upstream-downstream relation, this control strategy can adapt to changes in the network topology and supports a flexible plug-and-play operation.

\section{Coalition Formation}

Upon initialization, the coalition formation scheme starts with a single coalition consisting of every PV inverter. It then periodically alters the set of coalitions to:
\begin{itemize}
    \item Separate the over- and under-voltage violations and assign them to different coalitions to prevent conflicts between PV inverters;
    \item Allocate sufficient voltage regulation capacity to each coalition;
    \item Maintain the connectivity of the communication network within each coalition.
\end{itemize}
Each time the coalitions are updated, the PV inverters first gather necessary data. Based on this information, they evaluate the current coalition status and determine their actions.

\subsection{Information Gathering}
Before updating the coalitions, each inverter $i$, ${i \in \mathcal{G}_a}$ needs to gather the following data:
\begin{itemize}
    \item Local data: the local voltage magnitude $V_i$ and utilization ratio $u_i$;
    \item Neighbor data: the voltage magnitude $V_j$ and utilization ratio $u_j$ of neighbors $j$, ${\forall{j} \in \mathcal{N}_i}$;
    \item Coalition data: the assessed coalition maximum voltage $V_i^{\mathrm{max}}$ and minimum voltage $V_i^{\mathrm{min}}$ of $\mathcal{G}_a$. 
\end{itemize}
Algorithm 1 shows how inverters obtain the coalition data through neighboring communication. Each inverter measures its local voltage $V_i$ and uses it to initialize its estimate of the coalition maximum voltage $V_i^{\mathrm{max}}$ and minimum voltage $V_i^{\mathrm{min}}$. At each iteration within the time window ${[T-\Delta T, \: T]}$, the inverter communicates with its neighbors within the same coalition to exchange their estimates. As the iterations progress $V_i^{\mathrm{max}}$ and $V_i^{\mathrm{min}}$ are updated based on the max and min-consensus algorithm. Since the communication network is connected, in a coalition with $|\mathcal{V}_a|$ inverters, after a maximum ${|\mathcal{V}_a|-1}$ iterations \cite{maxconsensus}, $V_i^{\mathrm{max}}$ and $V_i^{\mathrm{min}}$ converge to the coalition maximum voltage $V_a^{\mathrm{max}}$ and minimum voltage $V_a^{\mathrm{min}}$. 

\begin{algorithm}
\caption{Algorithm for assessing coalition voltage status}
\textbf{Initialization}: ${V_i^{\mathrm{max}}(0) = V_i}$ and ${V_i^{\mathrm{min}}(0) = V_i}$ \\
\textbf{Iteration}: At the $k$th iteration: \\
Step 1: Sends $V_i^{\mathrm{max}}(k)$ and $V_i^{\mathrm{min}}(k)$ to neighbors ${j \in \mathcal{N}_{i}^a}$\\
Step 2: Receives $V_j^{\mathrm{max}}(k)$ and $V_j^{\mathrm{min}}(k)$ from neighbors ${j \in \mathcal{N}_{i}^a}$\\
Step 3: Updates these two variables as follows:
\begin{equation}
  \begin{cases}
    V_i^{\mathrm{max}}(k+1) = \max (V_i^{\mathrm{max}}(k), V_{j \in \mathcal{N}_i^a}^{\mathrm{max}}(k)) \\
    V_i^{\mathrm{min}}(k+1) = \min (V_i^{\mathrm{min}}(k), V_{j \in \mathcal{N}_i^a}^{\mathrm{min}}(k))
  \end{cases}   
\end{equation}
where ${\mathcal{N}^a_i = \{j \in \mathcal{V}_a: (i, j) \in \mathcal{E}_a\}}$ is the set of inverter $i$'s neighbors within the same coalition as $i$. 
\end{algorithm}

\begin{figure*}[bhtp!]
\centering
\includegraphics[width=6.25in]{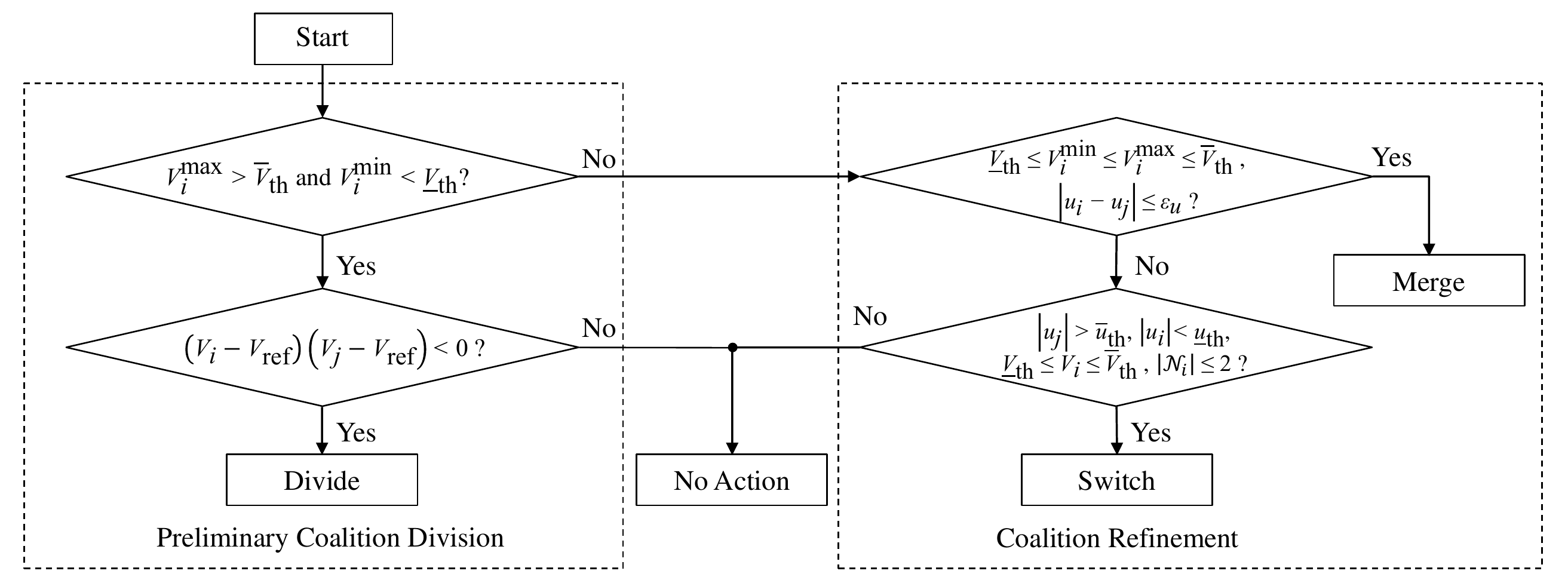}
\caption{Decision logic for updating the coalitions.}
\label{GroupFormulation}
\end{figure*}

\begin{figure*}[bhtp!]
\centering
\includegraphics[width=5.25in]{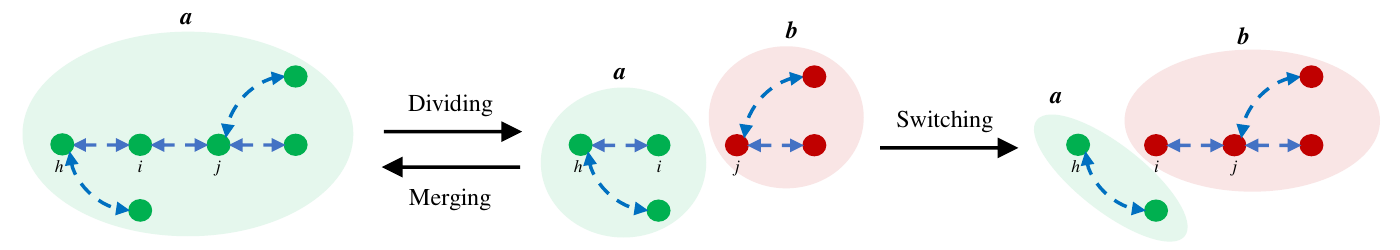}
\caption{Coalition transformation under different updating actions.}
\label{GroupFormulationE}
\end{figure*} 

\subsection{Coalition Updating}
Having gathered these data, each PV inverter follows the decision logic of Fig.~\ref{GroupFormulation} to determine if the coalition it currently belongs to needs to be divided, refined or left intact. Fig.~\ref{GroupFormulationE} shows the transformation of the coalition under different updating actions. 

\subsubsection{Dividing}
PV inverter $i$, ${i \in \mathcal{G}_a}$ first compares the coalition maximum voltage $V_i^{\mathrm{max}}$ and minimum voltage $V_i^{\mathrm{min}}$ with the upper and lower voltage thresholds, $\bar{V}_\mathrm{th}$ and $\ubar{V}_\mathrm{th}$, to determine whether over and under-voltage problems coexist in the coalition. If this is the case, inverter $i$ will decide to separate from its neighbor $j$, ${j \in \mathcal{G}_a}$ if its local voltage magnitude $V_i$ is lower than the voltage reference value $V_{\mathrm{ref}}$ while its neighbor $j$'s is higher, or vice-versa. If inverter $i$ decides to separate from $j$, $j$ will draw the same conclusion. They will therefore ignore the information they receive from each other in the calculation of the utilization ratio. As they no longer cooperate, the coalition $ \mathcal{G}_a$ is split. For the coalition communication network, this action is equivalent to the removal of the edge between nodes $i$ and $j$. Since the communication network of a typical distribution network has a tree topology, the removal of edges decomposes the original communication network into disjoint sub-networks. Each of these smaller networks still forms a connected graph with tree topology and supports the interactions of PV inverters in the newly formed coalitions. This voltage-based coalition division rule ensures that the PV inverters experiencing an upper voltage limit violation can be separated from those experiencing a lower voltage limit violation. Moreover, the PV inverters whose voltage is closer to the upper voltage limit are generally grouped with the former, while the PV inverters whose voltage is closer to the lower voltage limit are grouped with the latter.

\subsubsection{Merging}
If no further division is needed for a coalition, the PV inverters execute the capacity-based coalition refinement rules to improve the distribution of voltage regulation capacity among coalitions. For example, when both the maximum and minimum voltages of a coalition $\mathcal{G}_a$ have returned to the threshold range ${[\ubar{V}_\mathrm{th}, \: \bar{V}_\mathrm{th}]}$, PV inverter $i$, ${i \in \mathcal{G}_a}$ can reasonably assume the over/under-voltage problem that existed before has been solved and the coalition can merge with others to share their voltage regulation burden. Therefore, if an inverter had separated from a neighbor, it checks the utilization ratio $u_j$ of that neighbor $j$, ${j \in \mathcal{G}_b}$. If the difference between $u_j$ and its own utilization ratio $u_i$ is smaller than a threshold $\varepsilon_u$, then inverter $i$ determines that it is safe to merge $\mathcal{G}_a$ with $\mathcal{G}_b$. Since this merger does not cause a large change to the reactive power output of the PV inverters currently in $\mathcal{G}_a$, this action would not create voltage problems for them. PV inverter $i$ then reconnects with $j$, and thus $\mathcal{G}_a$ merges with $\mathcal{G}_b$. This action reconnects node $i$ and $j$ in the communication network, to form a larger connected network that supports the communications within the newly merged coalition. 

\subsubsection{Switching}
A coalition $\mathcal{G}_a$ cannot merge as a whole with another coalition $\mathcal{G}_b$ if its voltage problems persist or the difference in their utilization ratio is too large. However, some inverters can switch from coalition $\mathcal{G}_a$ to $\mathcal{G}_b$ when $\mathcal{G}_b$ suffers from a lack of voltage regulation capacity. Assuming the utilization ratio of $j$, ${j \in \mathcal{G}_b}$ exceeds an upper threshold $\bar{u}_\mathrm{th}$, that is, almost all the voltage regulation capacity has been used up in coalition $\mathcal{G}_b$. When $j$'s neighbor $i$, ${i \in \mathcal{G}_a}$ detects this situation, it can switch to $\mathcal{G}_b$ and thus bring extra regulation capacity to $\mathcal{G}_b$ under two conditions: 1) The coalition $\mathcal{G}_a$ has extra capacity, i.e. ${|u_i| < \ubar{u}_\mathrm{th}}$. 2) This switch does not bring a new voltage problem to coalition $\mathcal{G}_b$, i.e. ${\ubar{V}_\mathrm{th} \le V_i \le \bar{V}_\mathrm{th}}$. Moreover, to preserve the connectivity of a coalition's communication network, this switching action is limited to PV inverters with no more than two neighbors. For example, in Fig.~\ref{GroupFormulationE}, PV inverter $i$ can switch to $\mathcal{G}_b$ when necessary, while $j$ is not allowed to switch coalitions because this would split the communication network of $\mathcal{G}_b$. For the communication network, the impact of a switch is the same as removing the edge between node $i$ and its neighbor ${h \in \mathcal{G}_a}$ while reconnecting node $i$ with its other neighbor ${j \in \mathcal{G}_b}$. Neither action would impact the connectivity or the tree topology of the newly formed coalitions' communication networks. 

If the above conditions are not satisfied, the PV inverter takes no action in this coalition formation time interval.

\subsection{Discussion}
The actions of individual PV inverters periodically update the coalitions to solve potential conflict between voltage regulation and reactive power sharing. Compared with other strategies, the main characteristics of this coalition formation scheme are as follows:

\begin{itemize}
    \item This scheme provides a simple, clear and uniform decision logic to each PV inverter. The execution of this logic immediately results in a unique set of coalitions. Other dynamic coalition formation schemes involve several rounds of negotiation between potential partners \cite{Schedule, Trading}, or require a coordinating agent to construct the coalition gradually by contacting every possible member sequentially \cite{Restoration}.
    \item This scheme allows the inverters to flexibly organize themselves using dividing, merging and switching actions in response to real-time network operating condition. In contrast, centralized network decomposition-based strategies \cite{Partition1, Partition2} use the voltage sensitivity matrix to divide the network and the regulation resources. These approaches give inverters less flexibility to determine the scope of their cooperation.
\end{itemize}

\section{Coordination within a Coalition}

Once a coalition has been formed, the PV inverters within that coalition autonomously select one member as their leader. A feedback-based leader-follower consensus algorithm then coordinates the real-time control of the inverters.  The leader adjusts its output to eliminate the voltage violation. The followers, i.e. the other inverters in the coalition, adjust their outputs correspondingly to share the regulation burden.

\subsection{Role Identification}

In existing leader-follower consensus control strategies, the system operator usually predetermines which inverter is most likely to experience the largest voltage deviation and selects it as the leader for the entire control process \cite{Consensus1, Consensus2, Consensus3}. On the other hand, under our proposed control, the PV inverters  first determine the largest voltage deviation within their coalition and then identify their roles accordingly.
As in the coalition maximum voltage assessment process described in Algorithm 1, during a time window ${[T,\: T+\Delta T]}$ after the coalition formation, PV inverters exchange their local voltage deviations ${\Delta V_i = |V_i - V_{\mathrm{ref}}|}$ with their neighbors and update their estimation of the coalition maximum voltage deviation $\Delta V_i^{\mathrm{max}}$. The PV inverter whose voltage deviation is the largest assumes the role of leader, while the rest become followers. 

\subsection{Utilization Ratio Adjustment}
\subsubsection{Feedback Control}
After role identification, the leader inverter of a coalition $\mathcal{G}_a$ takes its local voltage magnitude $V_l$ as the feedback signal and adjusts its utilization ratio $u_a$ at each iteration $k$ as follows
\begin{equation} \label{eq: uref}
    u_a(k+1) = \ubar{\lambda}(k) - \bar{\lambda}(k)\\
\end{equation}
\begin{equation} \label{eq: lam1}
    \bar{\lambda}(k+1) = [\bar{\lambda}(k)+\alpha(V_l(k) - \bar{V})]^{+}
\end{equation}
\begin{equation} \label{eq: lam2}
    \ubar{\lambda}(k+1) = [\ubar{\lambda}(k)+\alpha(\ubar{V} - V_l(k))]^{+}
\end{equation} 
where $\bar{\lambda}$ and $\ubar{\lambda}$ are control states. $\alpha$ is the step-size parameter. $[*]^{+}$ denotes the projection operator onto ${[0, +\infty)}$. This calculation is similar to an integral controller with a dead-band. When the local voltage magnitude exceeds the upper voltage limit, the leader inverter reduces its utilization ratio $u_a$ and hence its reactive power output. On the other hand, if the local voltage magnitude is too low, it increases its utilization ratio to support the voltage.

\subsubsection{Leader-follower Consensus Algorithm}
The utilization ratio determined by the leader inverter is then communicated within coalition $\mathcal{G}_a$ according to the logic shown in Algorithm 2. At each iteration, inverter $i$ sends its utilization ratio to its neighbors in the same coalition and receives their utilization ratios. It then updates its utilization ratio to be the average of its own and its neighbors' utilization ratio. Since the communication network for the coalition is connected, a consensus on the utilization ratio $u_a$ will be reached globally and exponentially~\cite{Consensus}.

\begin{algorithm}
\caption{Algorithm for adjusting the utilization ratio}
\textbf{Iteration}: At the $k$th iteration: \\
 Step 1: Send $u_i(k)$ to neighbors ${j \in \mathcal{N}^a_{i}}$\\
 Step 2: Receive $u _j(k)$ from neighbors ${j \in \mathcal{N}^a_{i}}$\\
 Step 3: Update the utilization ratio: 
\begin{equation}  \label{eq: con}
    u_i(k+1) = \frac{1}{1 + |\mathcal{N}^a_{i}|} \left [u_i(k) + \sum_{j \in \mathcal{N}^a_{i}} u_j(k)\right ].
\end{equation}
\end{algorithm}

\subsubsection{Reactive Power Output}
The maximum available reactive power capacity of PV inverter $i$ is constrained by its rated capacity $S_{i}^{r}$ and real-time active power output $P_{i}$:
\begin{equation}
     Q_{i}^{\mathrm{max}} = \sqrt{{S_{i}^{r}}^{2} - P_{i}^{2}}.
\end{equation}
Since PV inverters are typically oversized, they can provide reactive power compensation even when they generate their rated active power output $P_{i}^{r}$,
\begin{equation}
     S_{i}^{r} = (1+\beta)P_{i}^{r}
\end{equation}
where $\beta$ is the over-sizing percentage. 
Based on the utilization ratio, the reactive power contribution from PV inverter $i$ for voltage regulation is
\begin{equation}
    Q_{i}(k) = \left [u_i(k) \right]_{-1}^{1} Q_{i}^{\mathrm{max}}
\end{equation}
where $[*]_{-1}^{1}$ denotes the projection operator onto the utilization ratio limit set ${[-1, 1]}$.

\subsection{Algorithm Analysis}
This subsection demonstrates the analytical connections between the feedback-based leader-follower consensus algorithm and distributed optimization algorithms.

Within a coalition, the voltage regulation can be formulated as the following optimization problem:
\begin{equation} \label{eq: Optimization}
\begin{split}
\text{min} \quad & \sum_{i \in \mathcal{G}_a} \frac{1}{2}\gamma_i Q_i^2  \\
\text{s.t.} \quad & \ubar{V} \le V_l \le \bar{V} \\
& Q_i \in \mathcal{Q}_i, \: \forall{i \in \mathcal{G}_a}
\end{split}
\end{equation}
where ${\gamma_i > 0}$ is a penalty parameter associated with the reactive power output of inverter $i$ and ${\mathcal{Q}_i = [-Q_{i}^{\mathrm{max}},\: Q_{i}^{\mathrm{max}}]}$. The local voltage magnitude of the leader inverter can be approximated by linearizing the power-flow equation as:
\begin{equation}
    V_l \approx \sum_{i \in \mathcal{G}_a} \frac{\partial V_l}{\partial Q_i}Q_i + V_\mathrm{uncon}
\end{equation}
where ${\frac{\partial V_l}{\partial Q_i} > 0}$ is the sensitivity of the local voltage at the leader inverter to the reactive power output of inverter $i$. $V_\mathrm{uncon}$ is the uncontrollable part of the voltage magnitude, which depends on the load and active PV generation. 

Assuming problem~(\ref{eq: Optimization}) is feasible and the Slater condition is satisfied, i.e. there exist ${Q_i, \: \forall{i \in \mathcal{G}_a}}$ such that ${Q_i \in \mathcal{Q}_i}$ and ${\ubar{V} < V_l < \bar{V}}$. Given the strong convexity of the cost function, this problem has a unique optimal solution which can be obtained using a projected primal-dual dynamics feedback optimization algorithm \cite{Convergence1}:
\begin{equation} \label{eq: Qi}
    \Dot{Q}_i = \Pi_{\mathcal{T}_{\mathcal{Q}_i}^{Q_{i}}} \left[-\gamma_i Q_i + \frac{\partial V_l}{\partial Q_i} (\ubar{\lambda}-\bar{\lambda}) \right], \: \forall{i \in \mathcal{G}_a}
\end{equation}
\begin{equation} \label{eq: lambda1}
    \dot{\bar{\lambda}} = \Pi_{\mathcal{T}_{\mathbb{R}_{+}}^{\bar{\lambda}}}(V_l - \bar{V})
\end{equation} 
\begin{equation} \label{eq: lambda2}
    \dot{\ubar{\lambda}} = \Pi_{\mathcal{T}_{\mathbb{R}_{+}}^{\underline{\lambda}}}(\ubar{V}-V_l)
\end{equation} 
where ${\bar{\lambda}, \: \ubar{\lambda} \in \mathbb{R}_{+}}$ are the dual variables for the upper and lower voltage limit constraints in problem~(\ref{eq: Optimization}). Due to the special structure of the objective function in problem~(\ref{eq: Optimization}), as proved in \cite{Dopt1}, (\ref{eq: Qi}) can be simplified as
\begin{equation} \label{eq: Qis}
    Q_i = \left [\gamma_i^{-1} \frac{\partial V_l}{\partial Q_i} (\ubar{\lambda}-\bar{\lambda}) \right]_{-Q_{i}^{\mathrm{max}}}^{Q_{i}^{\mathrm{max}}}, \: \forall{i \in \mathcal{G}_a}.
\end{equation}
Assuming in problem~(\ref{eq: Optimization}), ${\gamma_i = (Q_{i}^{\mathrm{max}})^{-1}\frac{\partial V_l}{\partial Q_i}}$. (\ref{eq: Qis}) becomes
\begin{equation} \label{eq: ui}
    u_i = \frac{Q_i}{Q_{i}^{\mathrm{max}}} = \left[(\ubar{\lambda}-\bar{\lambda}) \right]_{-1}^{1}, \: \forall{i \in \mathcal{G}_a}.
\end{equation}

To execute this "gather and broadcast" algorithm under the neighboring communication constraint, different distributed algorithms have been proposed \cite{Convergence1, Dopt1}. The algorithm we implement is similar to the method in \cite{Convergence1}. The leader inverter (i.e., actuators who access sensor data in \cite{Convergence1}) measures its local voltage and updates the dual variables $\bar{\lambda}$, $\ubar{\lambda}$ according to (\ref{eq: lambda1}) and (\ref{eq: lambda2}). The discrete-time version of this process is (\ref{eq: lam1}) and (\ref{eq: lam2}). For the update of the utilization ratio, the leader inverter has direct access to ${(\ubar{\lambda}-\bar{\lambda})}$ and is thus able to adjust its utilization ratio according to (\ref{eq: ui}), i.e., (\ref{eq: uref}) in discrete-time. However, the followers (i.e., actuators who do not access sensor data in \cite{Convergence1}) rely on the peer-to-peer communication to pass the dual variables ${(\ubar{\lambda}-\bar{\lambda})}$ from the leader. Therefore, they keep a local copy ${(\ubar{\lambda}-\bar{\lambda})_i}$ of ${(\ubar{\lambda}-\bar{\lambda})}$ and update this local copy when information arrives from their neighbors. That is, the follower inverters preserve their own utilization ratio $u_i$ and update this value iteratively as described in Algorithm 2. It is proved in \cite{Convergence1} that this modified algorithm is guaranteed to converge to the optimal solution when the coupling in the communication graph is sufficiently strong. The feedback-based leader-follower consensus algorithm is thus able to eliminate upper or lower voltage limit violations within the coalition effectively and the utilization ratio converges to the optimal solution of problem~(\ref{eq: Optimization}). The detailed convergence proof can be found in \cite{Convergence1}.

\section{Implementation}
With their increased deployment, distributed energy resources (DERs) play a more active role in system operation. Various stakeholders have been developing standards to specify DER interconnection criteria and requirements (e.g., IEEE 1547-2018), as well as to guide DER interoperability and communication (e.g., IEEE 1815, IEEE 2030.5, IEC 61850). Accordingly, inverters are evolving towards more advanced, standardized and autonomous functionalities, which makes their cooperative control actions possible \cite{Smart2}. For example, a project by the Toronto and Region Conservation Authority (TRCA) demonstrated the practicality of organizing the smart inverters under a multi-agent peer-to-peer communication framework for distribution network voltage regulation \cite{Cooperative}. Our proposed voltage regulation strategy coordinates the smart PV inverters located on the low-voltage network under a similar framework. For large-scale deployment, considering there are abundant utility-owned voltage regulation devices on the primary side and the cost for communication, it is recommended to divide the distribution network and implement the strategy on each low-voltage network independently. An important obstacle is that these inverters are usually owned by utility customers. The utility would therefore not be able to coordinate their actions directly. To address this issue and encourage the participation of smart inverters in voltage regulation, incentive schemes \cite{Incentive} and alternative forms of inverter ownership \cite{Ownership} are being considered. Since the proposed strategy supports flexible plug-and-play operation, it can accommodate a variable number of inverters.

\section{Case Studies}

\subsection{Test System}
\begin{figure}[ht!]
\centering
\includegraphics[width=3.5in]{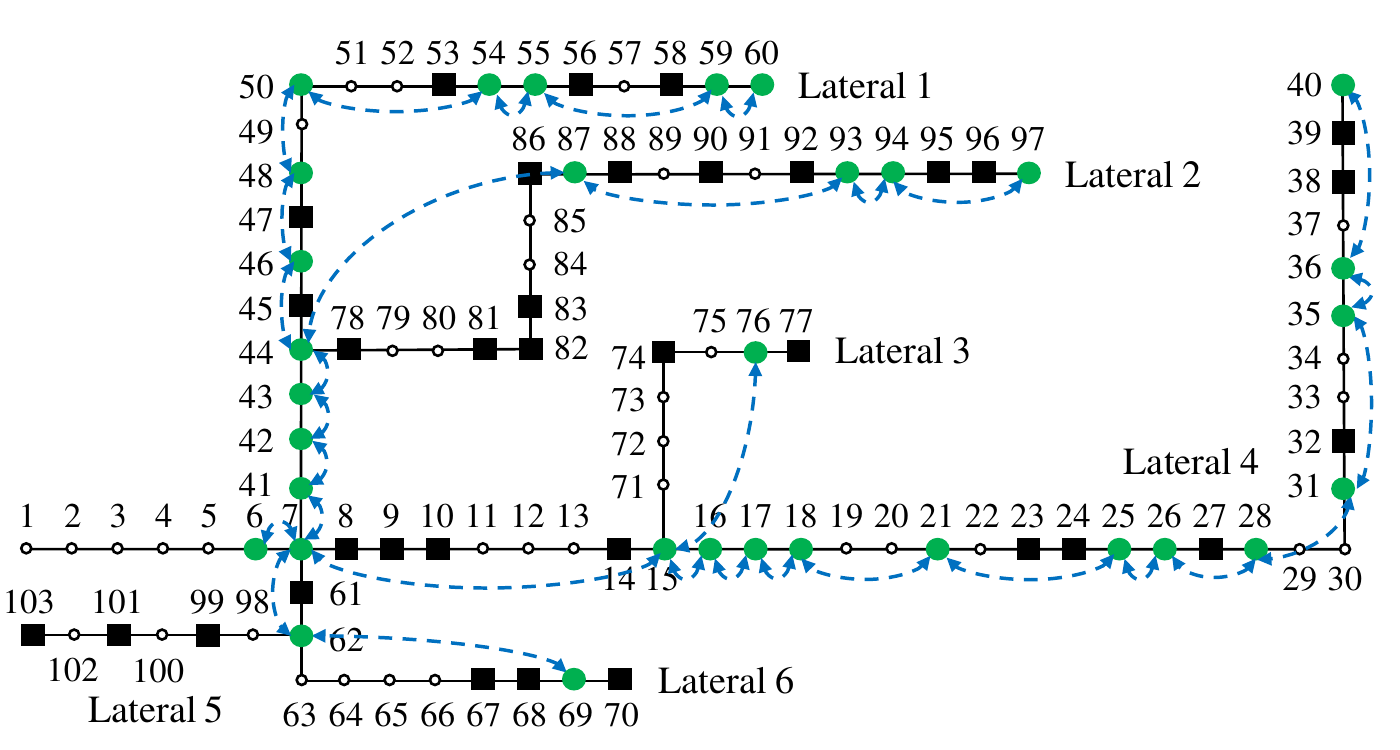}
\caption{Single-line diagram of the test system.}
\label{Network}
\end{figure}

\begin{figure}[ht!]
\centering
\includegraphics[width=3.5in]{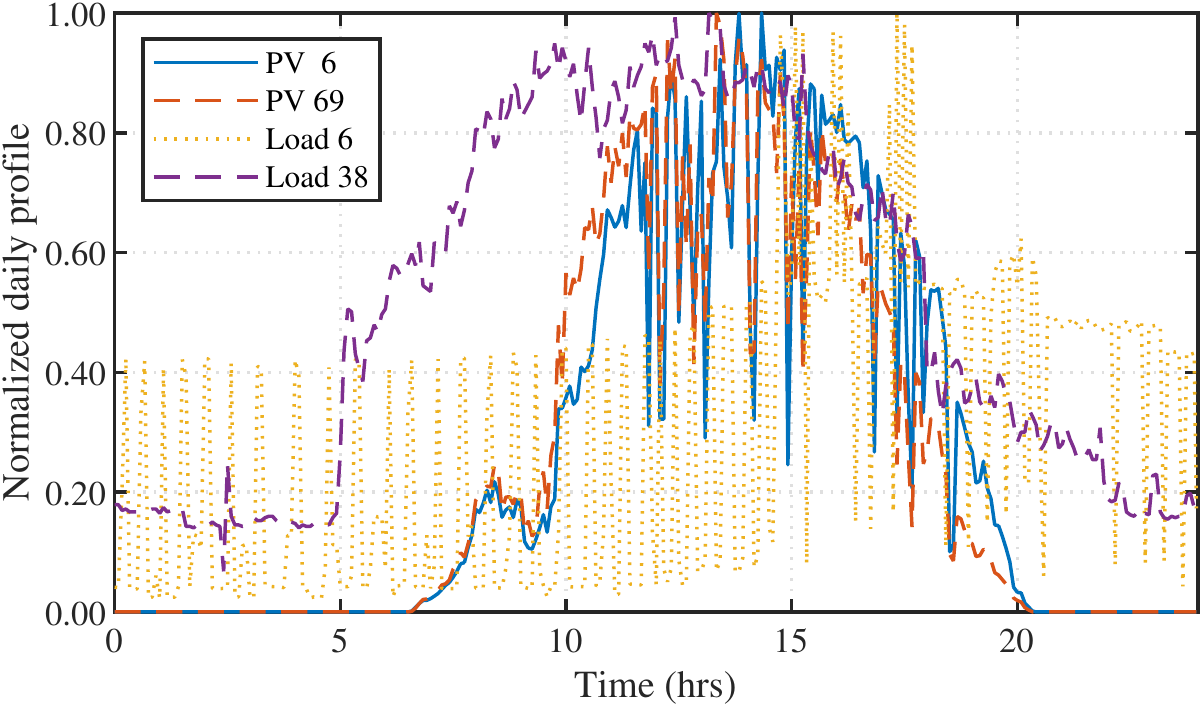}
\caption{Normalized daily profiles of the PV located at nodes 6 and 69 and the load located at nodes 6 and 38.}
\label{Prof}
\end{figure}

\begin{figure}[ht!]
\centering
\includegraphics[width=3.5in]{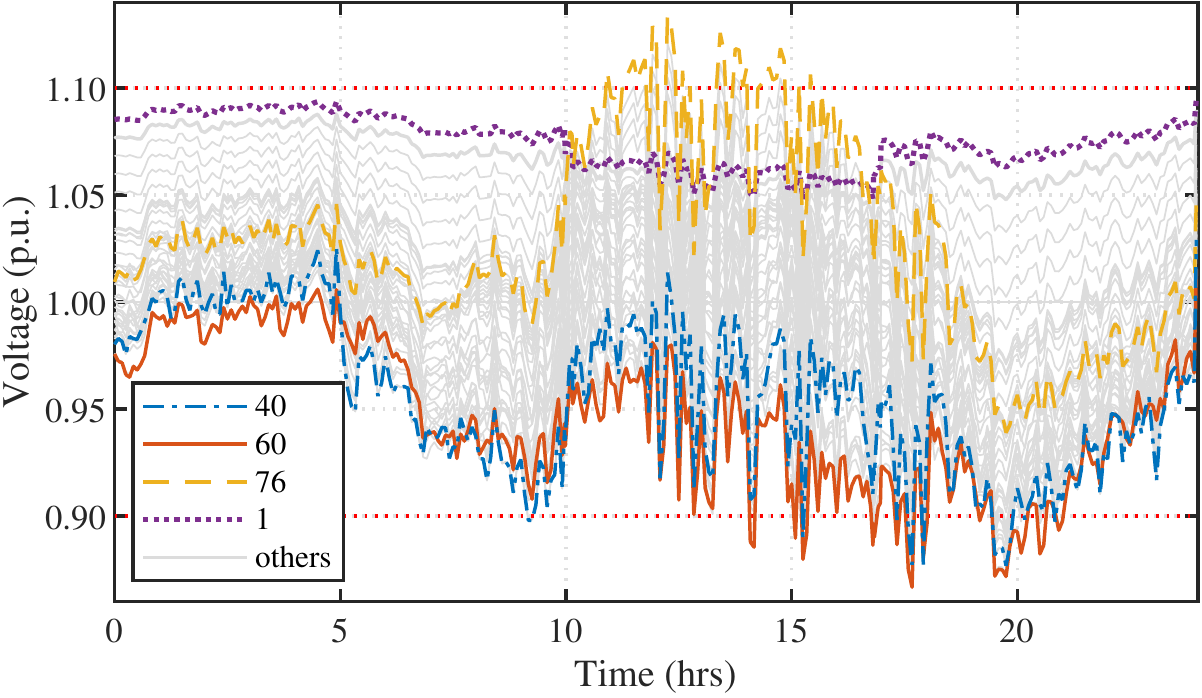}
\caption{Voltage profiles of the network without PV reactive power compensation over a 24-hour period.}
\label{NoControl}
\end{figure}

Fig.~\ref{Network} shows the single-line diagram of the balanced 230/400V 103-node radial distribution network used to demonstrate the performance of the proposed control strategy. The squares, circles and dashed lines represent the load nodes, PV nodes and the communication links, respectively. This network was extended based on the topology of a real semi-urban feeder in Flanders, Belgium. Its detailed network parameters can be found in \cite{Network}. To assess the effectiveness of the proposed strategy in a network where the load and PV generation are very unevenly distributed, we connected four heavy loads (a school, a bank, a grocery store and an office) at nodes 38, 56, 58, 96. Two small PV farms are located at nodes 69 and 76. Sixty houses are connected to the network, 30 of which are equipped with PV inverters. The 1-minute-resolution daily profiles for the PV generation and residential consumption were constructed based on the Pecan Street data set of June 16, 2014 \cite{Pecan}. The commercial load profiles are from \cite{EnerNOC}. Fig.~\ref{Prof} shows the normalized load and PV generation profiles. Table~\ref{tab: PV capacity} gives the capacity of these loads and DERs. According to the European Standard EN50160, the maximum allowable voltage deviation for this network is $\pm 10\%$. Fig.~\ref{NoControl} shows the voltage profiles of the distribution network when the only regulation device is the distribution transformer which connects node 1 to the substation. This transformer raises the system voltage during periods of heavy load and lowers the system voltage when the PV generation creates a high reverse power flow, i.e., from 10:00 to 16:30. However, during this period, even when the voltage at the end of laterals 1 and 4 hovers around the lower voltage limit, the voltage at the end of lateral 3 exceeds the upper voltage limit.

Although the analysis of the leader-follower consensus algorithm is built on a linearized power flow model, the power system is simulated using a full nonlinear AC power flow model based on MATPOWER \cite{MATPOWER}. The voltage regulation coalitions are reformed every 5 minutes and the sampling period of the local controllers is 200 milliseconds. To mitigate the impact of the volatility in PV generation, the voltage information gathered for coalition formation is its moving average value over a 15-minute rolling window. The control strategy regulation range $[\bar{V}, \ubar{V}]$ also reserves a small buffer. Moreover, as the coalition formation is taken as a preventive action, the threshold range $[\bar{V}_\mathrm{th}, \ubar{V}_\mathrm{th}]$ is set to be more conservative than the regulation range. Table~\ref{tab: control parameters} gives the detailed parameters of the proposed control strategy. This strategy's effectiveness is demonstrated through a 24-hour simulation. It is then compared with a simple localized controller and a state-of-the-art centralized network decomposition-based actuator organization scheme. The robustness and the generalizability of the proposed strategy are also assessed.

\begin{table}[ht!]
    \centering
    \caption{\textsc{Capacity of the Load and DERs}}
    \label{tab: PV capacity}
    \begin{tabular}{c c c c c}
        \toprule
        Type & Houses  &  School & Bank & Grocery\\
        \midrule
        Size (kW) & 2.2--14.5 & 68.4   & 31.4  &45.1 \\
        \midrule
         Type & Office & Residential PV & PV Farm 69 & PV Farm 76\\
        \midrule
        Size (kW) & 62.3 & 2.5--7.5 &45.5 &140.0 \\
        \bottomrule
    \end{tabular}
\end{table}

\begin{table}[ht!]
    \centering
    \caption{\textsc{Parameters of the Control Strategy}}
    \label{tab: control parameters}
\begin{tabular}{c c c c c c}
    \toprule
    $\bar{V}_\mathrm{th}$ & $\ubar{V}_\mathrm{th}$ & $\bar{V}$ & $\ubar{V}$ & $V_{\mathrm{ref}}$ \\
    \midrule
    1.05 p.u. & 0.95 p.u.&1.09 p.u. & 0.91 p.u. & 1.00 p.u.  \\
    \bottomrule
    $\varepsilon_u$ & $\bar{u}_\mathrm{th}$  & $\ubar{u}_\mathrm{th}$ & $\alpha$ & $\beta$ \\
    \midrule
    0.02 & 0.90 & 0.70 &  20 & 0.10 \\
    \bottomrule
\end{tabular}
\end{table}

\subsection{Performance Analysis}
\subsubsection{Convergence and Accuracy}
\begin{figure}[h!]
     \centering
     \begin{subfigure}[b]{0.5\textwidth}
         \centering
         \includegraphics[width=3.5in]{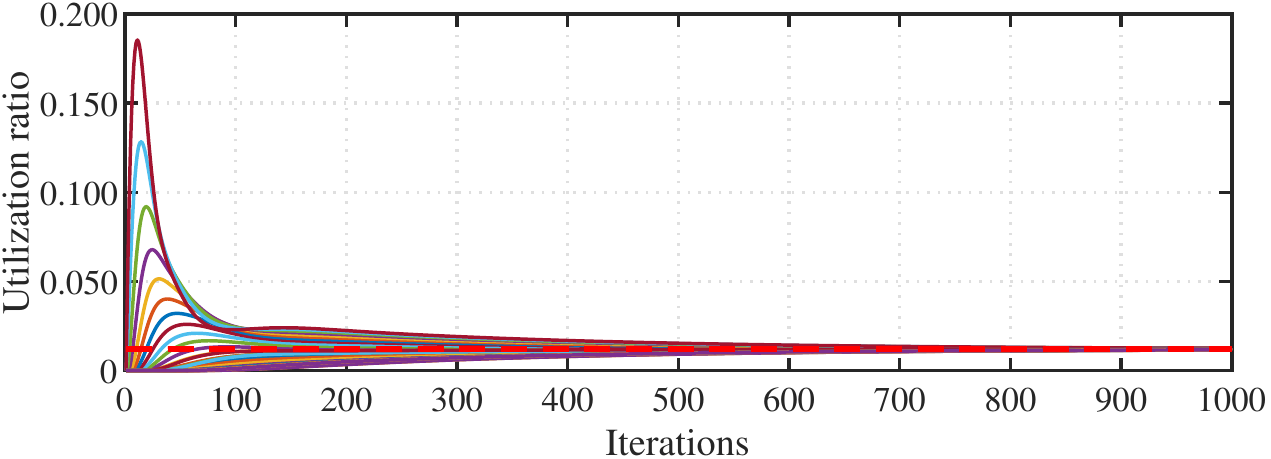}
         \caption{}
         \label{}
     \end{subfigure}
     \hfill
     \begin{subfigure}[b]{0.5\textwidth}
         \centering
         \includegraphics[width=3.5in]{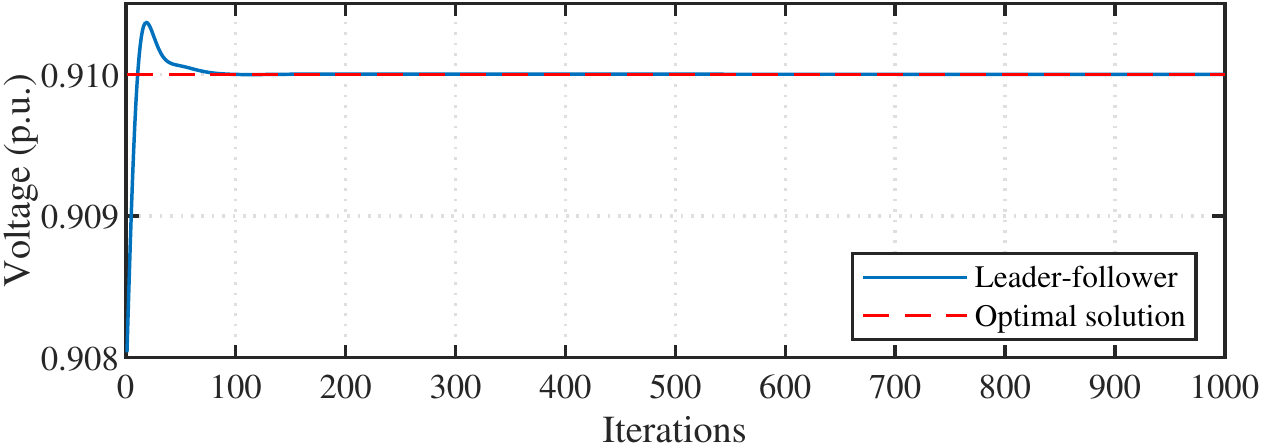}
         \caption{}
         \label{}
     \end{subfigure}
     \hfill
     \begin{subfigure}[b]{0.5\textwidth}
         \centering
         \includegraphics[width=3.5in]{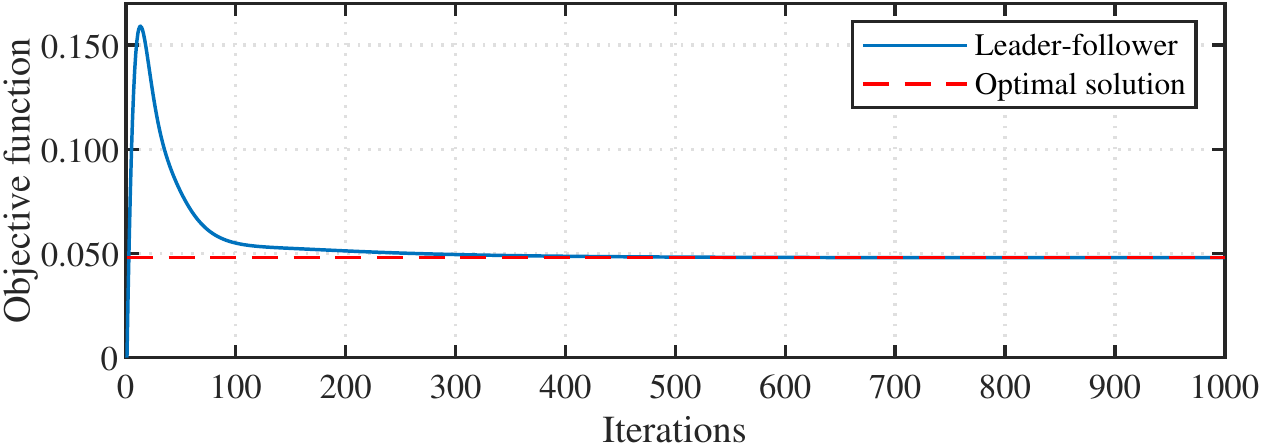}
         \caption{}
         \label{}
     \end{subfigure}
        \caption{Iteration process of the feedback-based leader-follower consensus algorithm. (a) Utilization ratios of each inverter, (b) Voltage magnitude at the leader inverter, (c) Objective function of problem~(\ref{eq: Optimization})}.
        \label{fig:Conv}
\end{figure}

Fig.~\ref{fig:Conv} shows the iteration process of the PV inverters assuming that the load and PV generation remain static at the values they have at 9:07. At this time, all the PV inverters belong to the same coalition and they work together in about 100 iterations, i.e. 20 seconds, to bring the voltage at node 40 back up to the prescribed lower voltage limit of 0.91 p.u. Meanwhile, the utilization ratio of each inverter gradually converges to the value determined by the leader inverter. The red dashed lines show the optimal solution of problem~(\ref{eq: Optimization}) where the voltage sensitivity value is calculated based on the Jacobian matrix, and the power flow equation constraints are modeled with the AC power flow model under Second-Order Cone Programming (SOCP) relaxation \cite{SOCP}. The leader-follower consensus algorithm drives the system to the optimal solution of problem~(\ref{eq: Optimization}), which confirms the analytical connection discussed in Section IV.

\subsubsection{Control over a 24-hour period}
\begin{figure}[htbp!]
\centering

\begin{subfigure}[b]{0.5\textwidth}
   \includegraphics[width=3.5in]{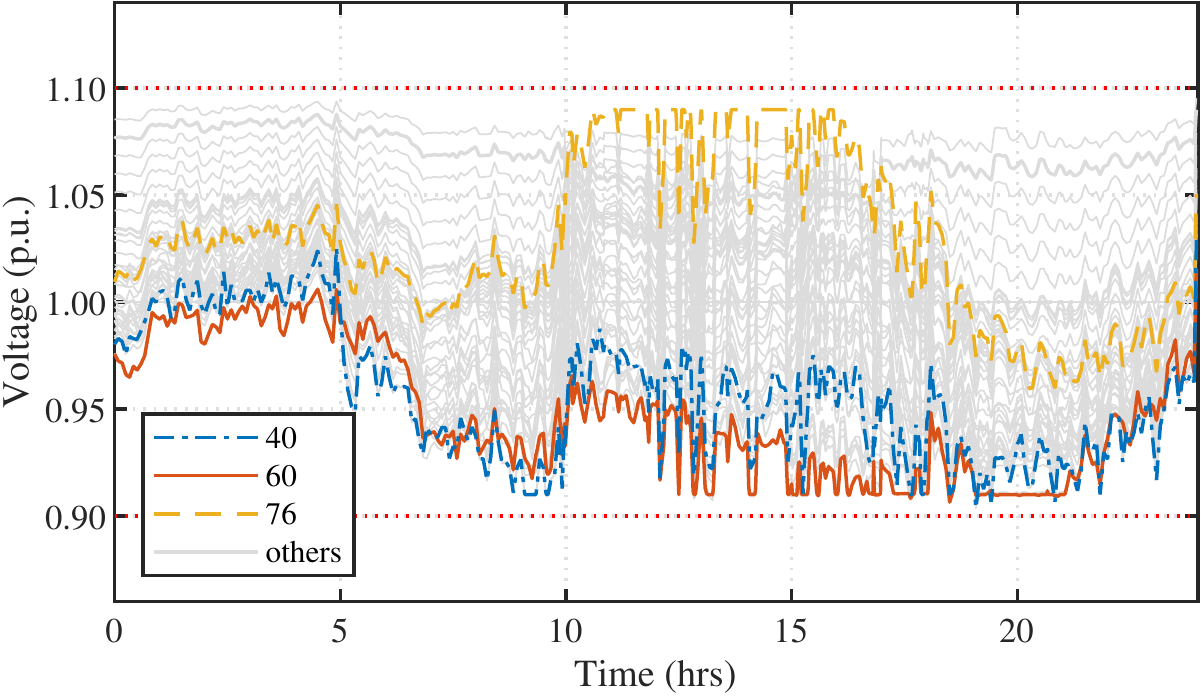}
   \caption{}
   \label{} 
\end{subfigure}

\begin{subfigure}[b]{0.5\textwidth}
   \includegraphics[width=3.5in]{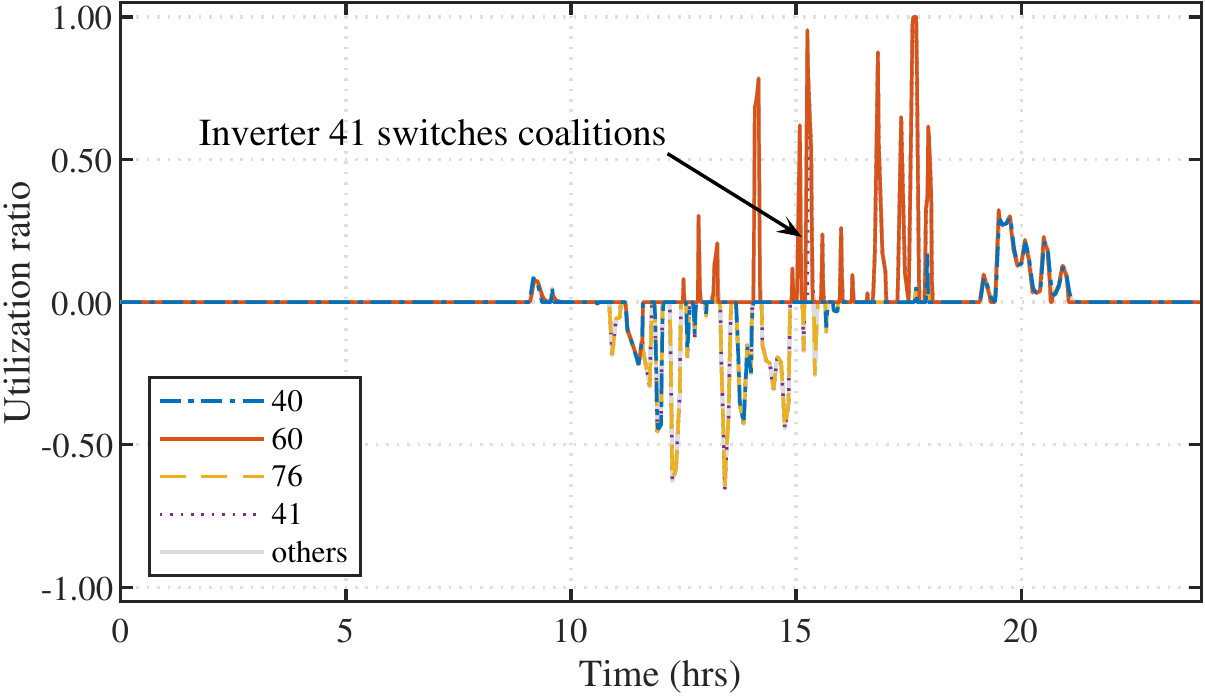}
   \caption{}
   \label{} 
\end{subfigure}

\caption{Voltage regulation results of the proposed control strategy. (a) 24-hour voltage profiles; (b) Utilization ratios.}
\label{WithControl}
\end{figure}

\begin{figure}[htbp!]
\centering
\includegraphics[width=3.5in]{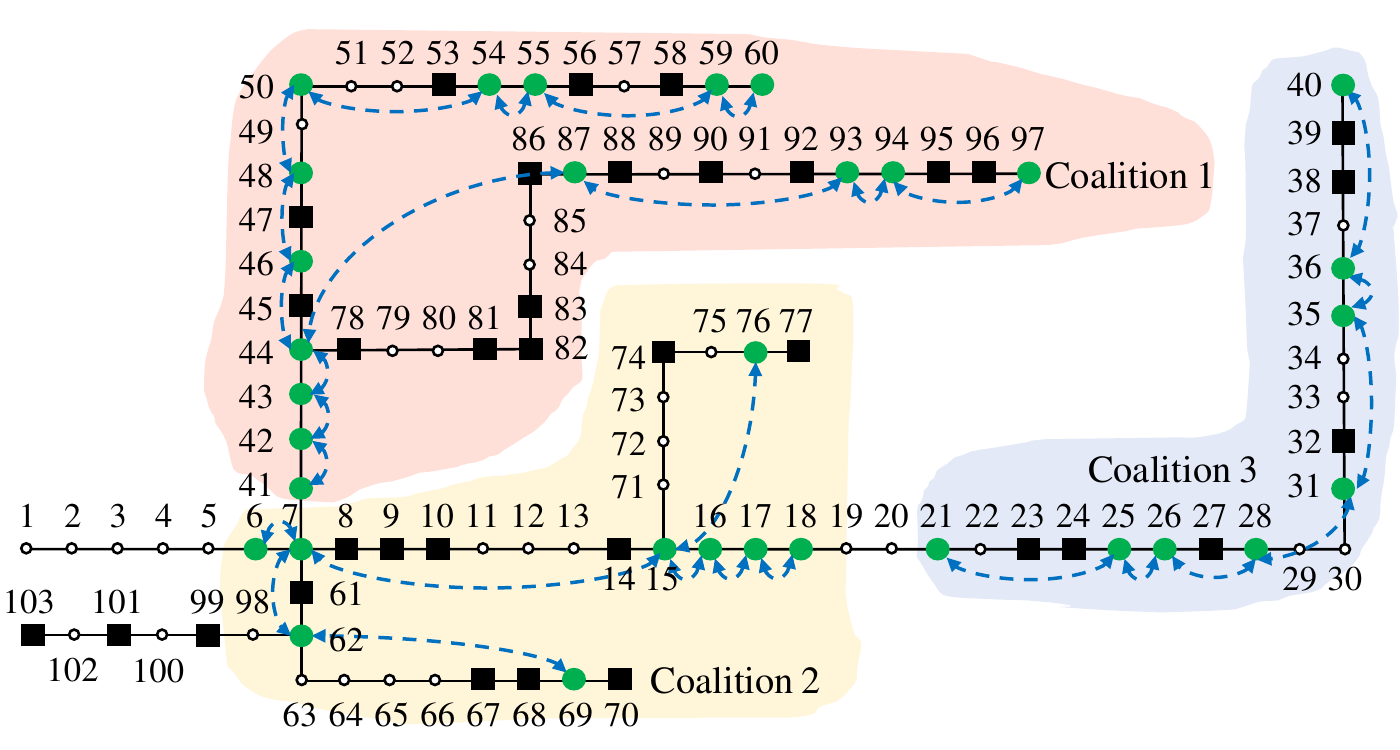}
\caption{Voltage regulation coalitions under the proposed strategy at 15:15.}
\label{GroupCon}
\end{figure}

Fig.~\ref{WithControl} illustrates the voltage regulation achieved by the proposed control strategy over the course of a day and demonstrates how the PV inverters are able to form into coalitions autonomously and solve the voltage violation problems cooperatively. Fig.~\ref{GroupCon} shows the coalitions that were formed at 15:15. Coalition 1 consists mainly of PV inverters on laterals 1 and 2, coalition 2 covers the inverters located on laterals 3, 6 and the first half of lateral 4. Inverters on the second half of lateral 4 form coalition 3. These coalitions evolve based on real-time voltage measurements. For example, coalition 3 reunites with coalition 2 to help solve its over-voltage problem at noon. However, this only happens when the voltages within coalition 3 are inside a safe range. Otherwise, coalition 3 separates from coalition 2 to prevent suffering from a lower voltage limit violation. In addition, coalitions can also be refined by the switching action of individual inverters. For instance, inverter 41 switches from coalition 2 to coalition 1 at 15:15 when coalition 1 lacks sufficient regulation capacity. Most of the time, coalitions 1, 2, 3 are led by inverters 60, 76, and 40, respectively.

\subsection{Comparison with Other Approaches}
\subsubsection{Local Control}
\begin{figure}[htbp!]
\centering

\begin{subfigure}[b]{0.5\textwidth}
   \includegraphics[width=3.5in]{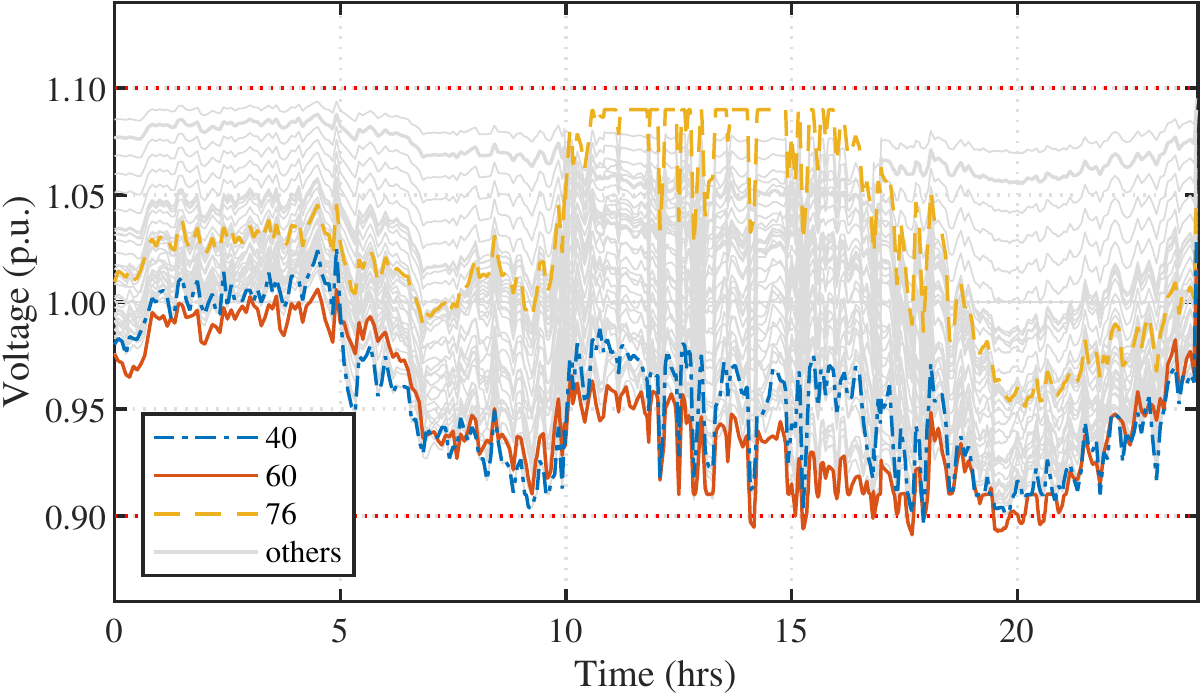}
   \caption{}
   \label{} 
\end{subfigure}

\begin{subfigure}[b]{0.5\textwidth}
   \includegraphics[width=3.5in]{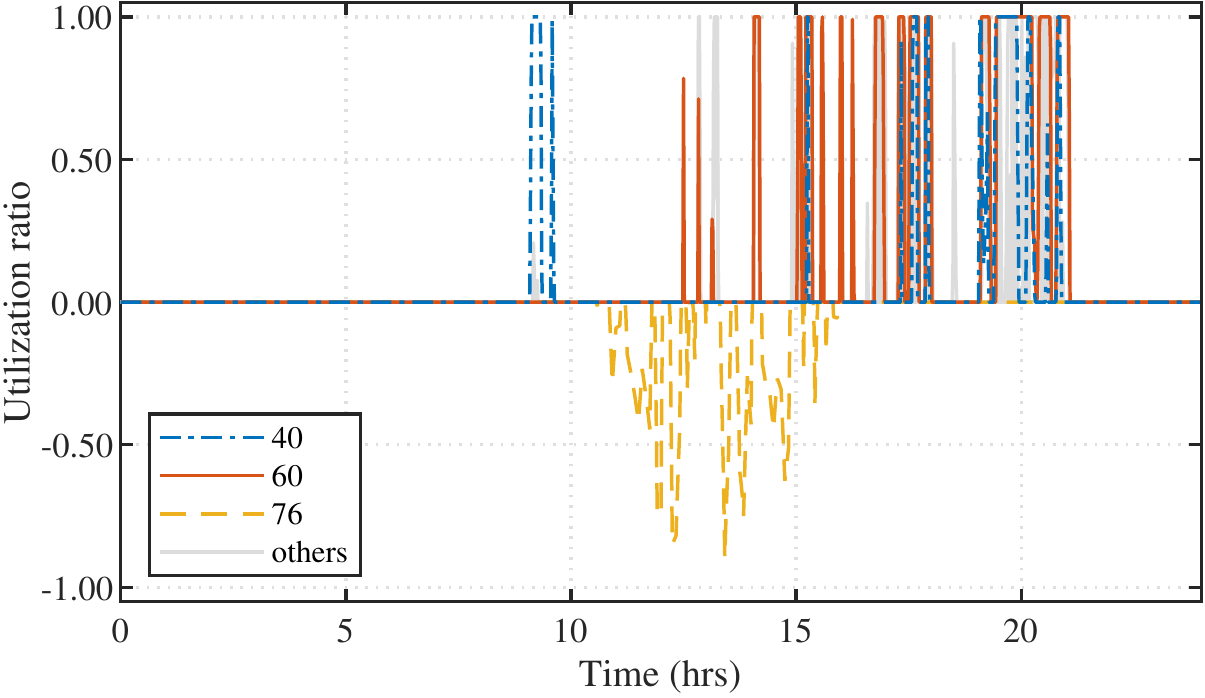}
   \caption{}
   \label{} 
\end{subfigure}

\caption{Voltage regulation results of the local control strategy. (a) 24-hour voltage profiles; (b) Utilization ratios.}
\label{LocalControl}
\end{figure}

When all the communication links are removed, the proposed strategy reduces to a simple local controller, the same as the one proposed in \cite{Integral}. In this case, each coalition contains a single inverter, which measures its local voltage and adjusts its reactive power output based on (\ref{eq: uref})-(\ref{eq: lam2}). Fig.~\ref{LocalControl} shows that this local control strategy is able to improve the voltage profile. However, lower voltage limit violations persist, mainly caused by the saturation of the PV inverters reactive power capacity. For example, during the time interval 15:13--15:19, only PV inverters 55, 59, 60 detect an under-voltage problem on lateral 1 and produce reactive power to solve it. However, due to their limited capacity and while their utilization ratios rise to the upper limit of 1.0, (i.e. they produce maximum reactive power), this voltage violation problem persists. To solve this problem, the active power generation from these PV inverters might need to be curtailed, which would lead to an economic loss for their owners. This problem does not occur with the proposed control strategy because coalition 1 in Fig.~\ref{GroupCon} involves more PV inverters that can help deal with these low voltages. Moreover, throughout the entire control process, only the PV inverters installed at the end of laterals actively participate in voltage regulation. Frequent actions and large reactive power output can affect the lifetime of these inverters \cite{Lifetime}. In contrast, the reactive power capacity of the rest of PV inverters is barely used. 

\subsubsection{Centralized Organization Scheme}
\begin{figure}[htbp!]
\centering

\begin{subfigure}[b]{0.5\textwidth}
   \includegraphics[width=3.5in]{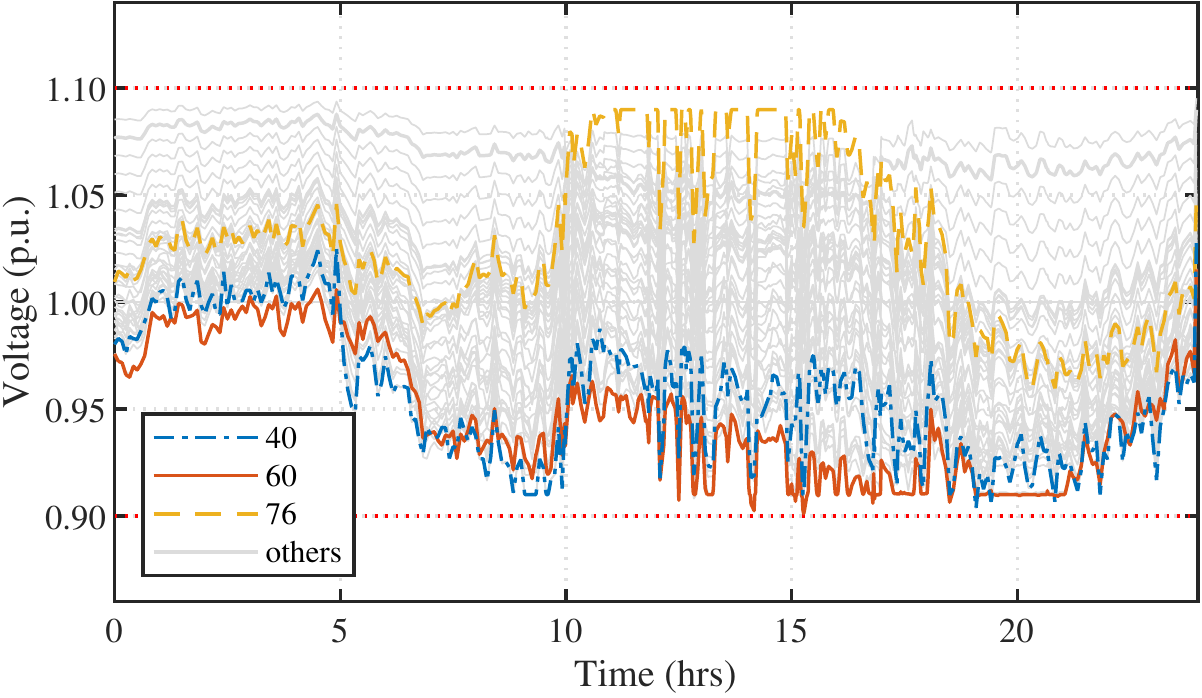}
   \caption{}
   \label{} 
\end{subfigure}

\begin{subfigure}[b]{0.5\textwidth}
   \includegraphics[width=3.5in]{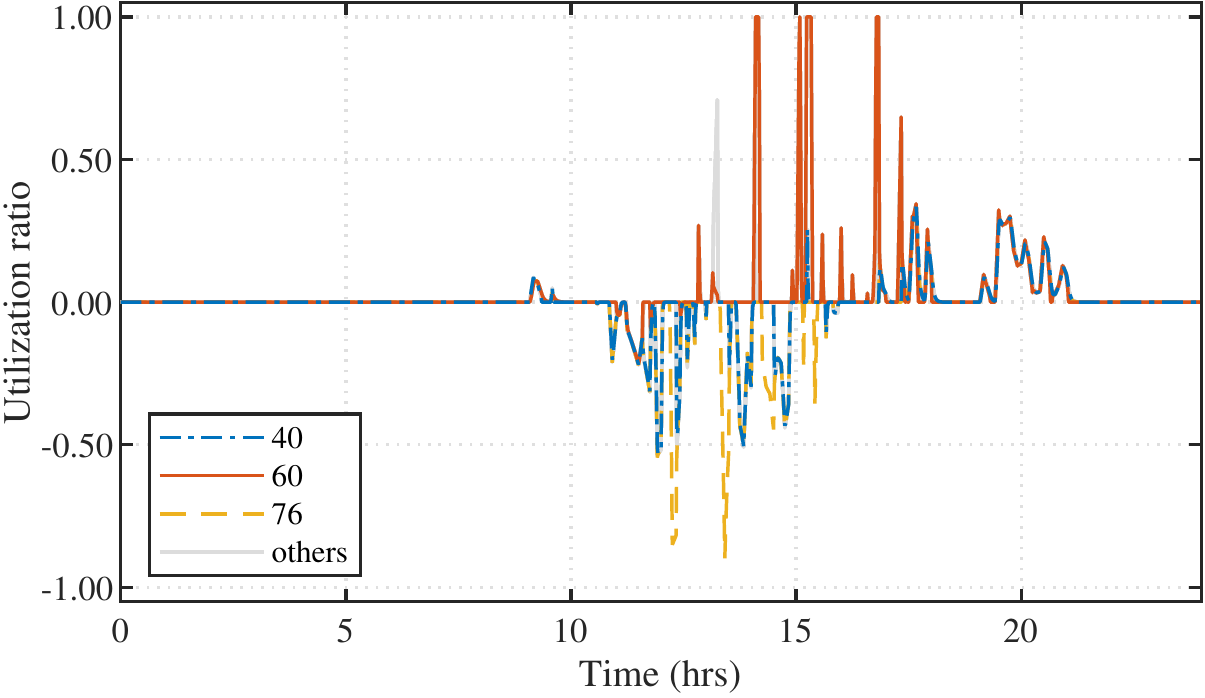}
   \caption{}
   \label{} 
\end{subfigure}

\caption{Voltage regulation results with the centralized organization scheme. (a) 24-hour voltage profiles; (b) Utilization ratios.}
\label{CenControl}
\end{figure}

\begin{figure}[htbp!]
\centering
\includegraphics[width=3.5in]{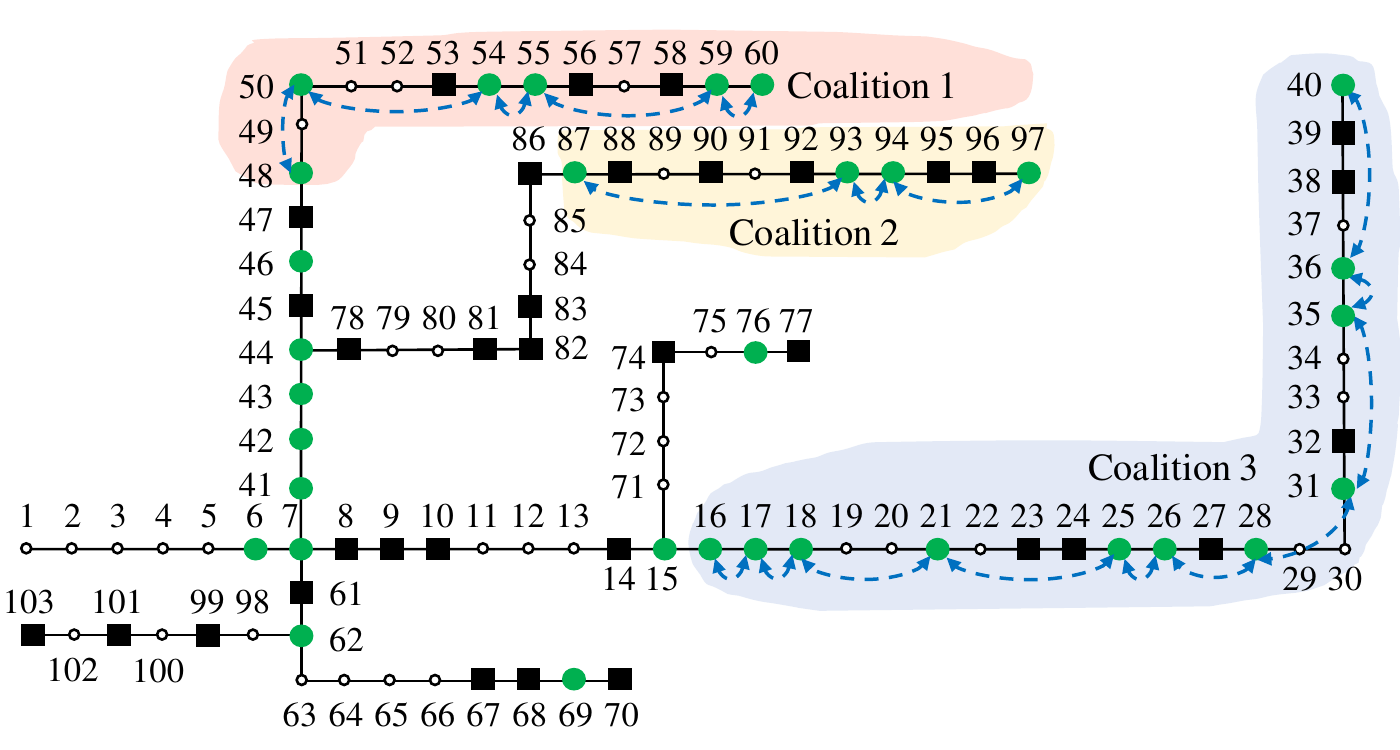}
\caption{Voltage regulation coalitions under centralized organization scheme at 15:15.}
\label{GroupCen}
\end{figure}

In the centralized actuator organization scheme proposed in \cite{Partition1}, a central agent derives the sensitivity matrix $A$ from the Jacobian matrix of the Newton-Raphson power flow solution:
\begin{equation} \label{eq: A}
    \begin{pmatrix}
      \Delta \theta\\ 
      \Delta V
    \end{pmatrix} = 
    \begin{pmatrix}
      A_{\theta P} & A_{\theta Q}\\ 
      A_{V P} & A_{V Q}
    \end{pmatrix}
    \begin{pmatrix}
      \Delta P\\ 
      \Delta Q
    \end{pmatrix}=A
    \begin{pmatrix}
      \Delta P\\ 
      \Delta Q
    \end{pmatrix}.
\end{equation}
This agent then applies epsilon decomposition to $A_{VQ}$:
\begin{equation} \label{eq: Epsi}
A_{VQ} = A_{VQ}' + \varepsilon B
\end{equation}
which decomposes the original voltage magnitude-reactive power sensitivity matrix $A_{VQ}$ into strong couplings $A_{VQ}'$ and weak couplings $B$. The weak couplings, i.e. any element less than $\varepsilon$, in $A_{VQ}$ are then set to 0. Correspondingly, the weakly coupled nodes in the network, and the inverters on these nodes, are partitioned into different voltage regulation zones. The central agent controls the partition result by updating the magnitude of $\varepsilon$ between 0 and 1. If $\varepsilon = 0$, all nodes are considered as strongly coupled and there is no partition, while if $\varepsilon = 1$, there is no cooperation. For comparison, we replace our coalition formation with this scheme. The central agent checks the network voltage magnitudes every 5 minutes. If a voltage threshold is violated, it initiates the partition and selects the smallest $\varepsilon$ to ensure that the inverters whose local voltage is higher than $\bar{V}_\mathrm{th}$ can be separated from those inverters whose local voltage is lower than $\ubar{V}_\mathrm{th}$. The leader-follower consensus algorithm still guides the within-coalition coordination. Fig.~\ref{CenControl} shows the corresponding voltage regulation results. Similar to our proposed scheme, this scheme is able to organize the inverters in different groups to solve the voltage violation problems. However, this coupling strength-based scheme is essentially different from our real-time voltage/capacity-based coalition formation scheme. Fig.~\ref{GroupCen} illustrates this difference. At 15:15, inverter 40 needs to be separated from inverter 76. Under our scheme, this separation is completed as inverter 18 separates from inverter 21 based on their voltage magnitudes. However, under the coupling strength-based scheme, the central agent finds the weakest coupling is between nodes 15 and 16 and sets the $\varepsilon$ value correspondingly to remove this coupling. However, couplings weaker than this $\varepsilon$, e.g. the coupling between 46 and 48, are also removed, which results in a smaller coalition 1 than ours. In general, with this network decomposition-based actuator organization scheme, inverters that are closer to the substation, and thus have a weaker coupling with their neighbors, are prone to be excluded from any cooperation.

\subsection{Robustness}
\subsubsection{Communication Latency}
\begin{figure}[htbp!]
\centering

\begin{subfigure}[b]{0.5\textwidth}
   \includegraphics[width=3.5in]{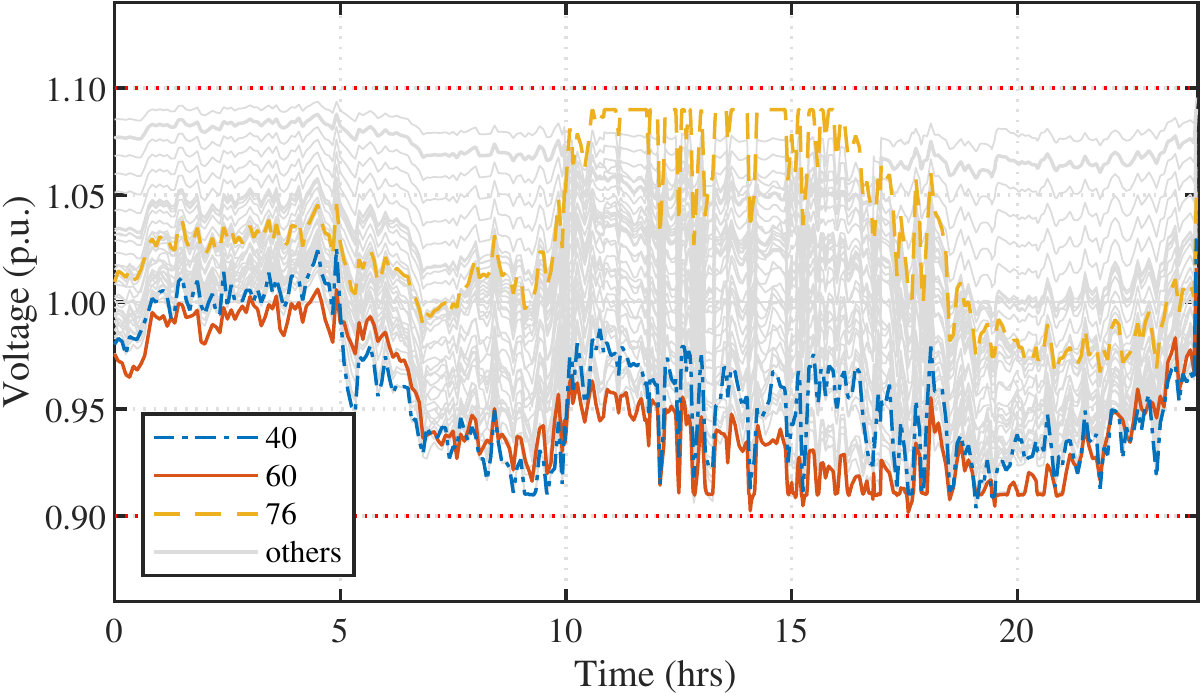}
   \caption{}
   \label{} 
\end{subfigure}

\begin{subfigure}[b]{0.5\textwidth}
   \includegraphics[width=3.5in]{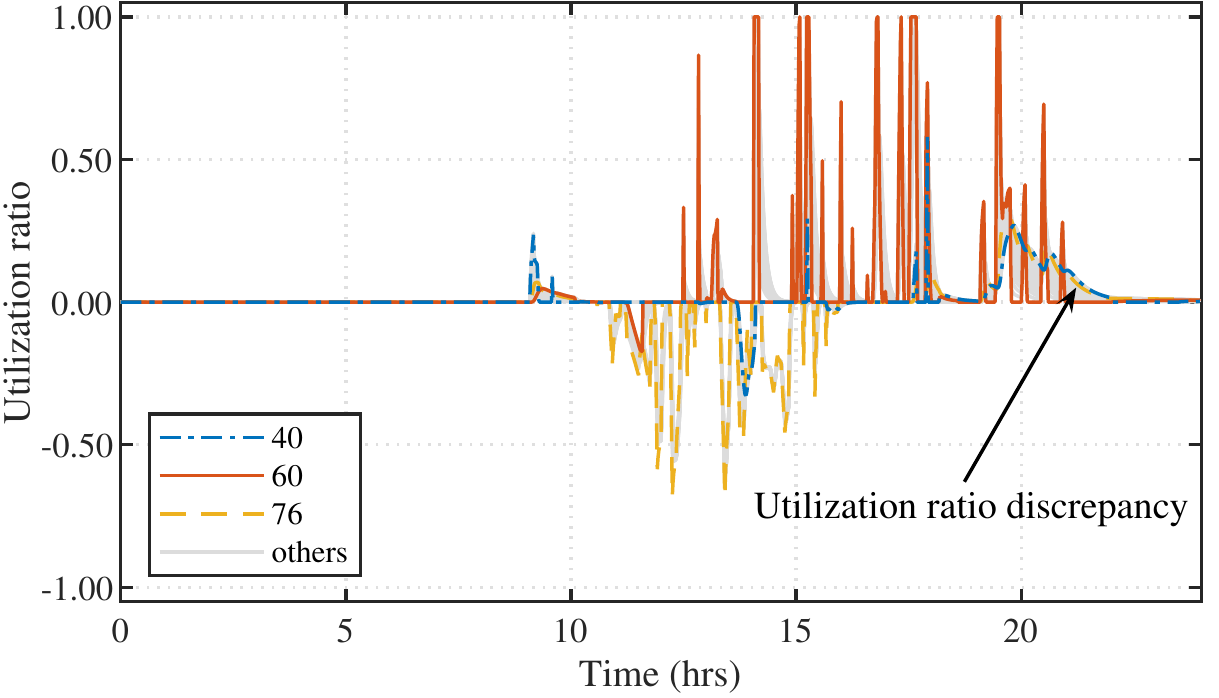}
   \caption{}
   \label{} 
\end{subfigure}

\caption{Voltage regulation results under communication latency. (a) 24-hour voltage profiles; (b) Utilization ratios.}
\label{LossControl}
\end{figure}

Communication latency is inevitable in real-world implementations. One way to handle the latency is to allocate a longer time for each iteration. To analyze the robustness of the proposed approach in handling latency, we increased the sampling period to 10 times its original value, i.e. 2000 milliseconds. Fig.~\ref{LossControl} shows the voltage regulation results. In this case, the coalition formation is barely impacted. This is because the coalitions are updated every 5 minutes, which allows the inverters sufficient time to assess the coalition voltage status and make decisions. On the other hand, it becomes more difficult for the inverters within the same coalition to fully communicate and reach consensus on their utilization ratios, particularly when the size of the coalition is large. For example, when all the inverters return to the same coalition after 20:00, discrepancies in the utilization ratio become obvious. However, the trends in these utilization ratios remain similar and the voltage regulation results are almost the same as in the normal case.

\subsubsection{Communication Failure}
\begin{figure}[htbp!]
\centering

\begin{subfigure}[b]{0.5\textwidth}
   \includegraphics[width=3.5in]{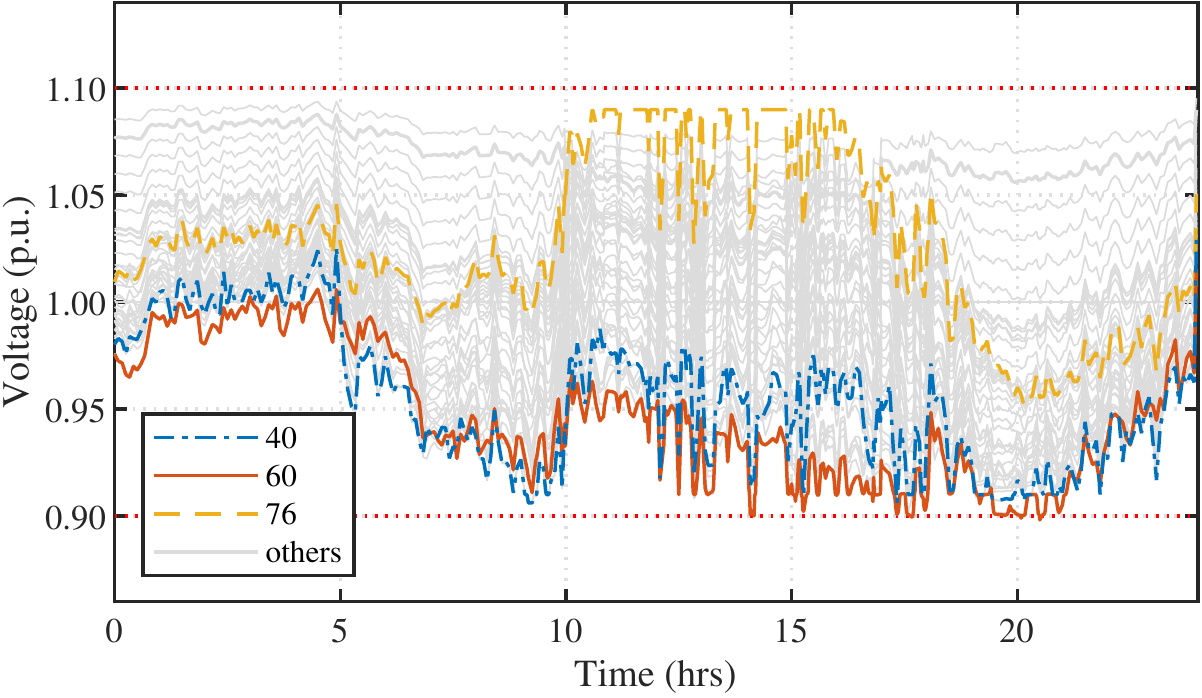}
   \caption{}
   \label{} 
\end{subfigure}

\begin{subfigure}[b]{0.5\textwidth}
   \includegraphics[width=3.5in]{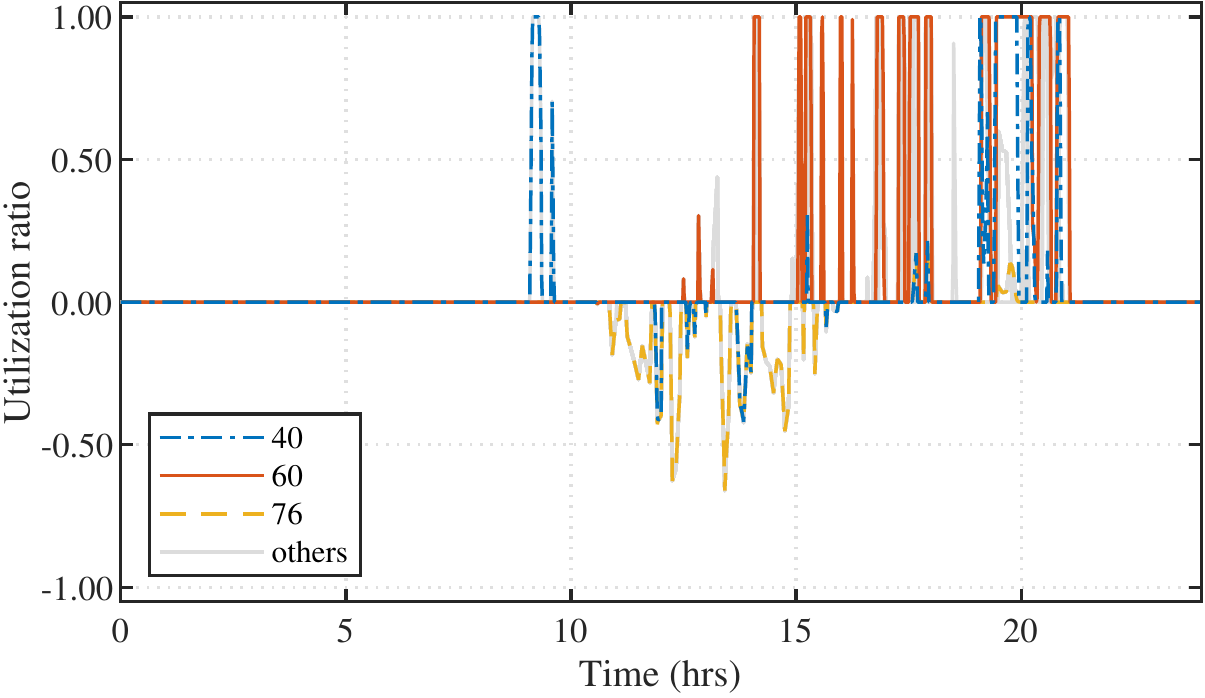}
   \caption{}
   \label{} 
\end{subfigure}

\caption{Voltage regulation results under communication failure. (a) 24-hour voltage profiles; (b) Utilization ratios.}
\label{CfControl}
\end{figure}

Communication failure is another problem that can deteriorate the effectiveness of the proposed strategy. For the robustness test, we assume that 10\% of the communication links fail at random every hour and these failure last for 15 minutes. For the inverters, a communication link failure is similar to a coalition division, as two neighboring inverters no longer cooperate. The difference is that the separated inverters are no longer able to merge or switch to each other's coalition since no information can be transmitted between them. Once the communication resumes, these actions are possible again. Fig.~\ref{CfControl} shows the voltage regulation results. In this case, as the unnecessary ``coalition division'' happens frequently, the scopes of cooperation are intermittently constrained. Therefore, the utilization ratios are quite different from the normal case and the coalition regulation capacity insufficiency problem can occur. However, as an adaptive coalition formation scheme, the proposed strategy is able to handle those undesirable separations when the communication recovers. The voltage magnitude can still be effectively constrained.

\subsection{Generalizability}
\begin{figure}[htbp!]
\centering
\includegraphics[width=3.5in]{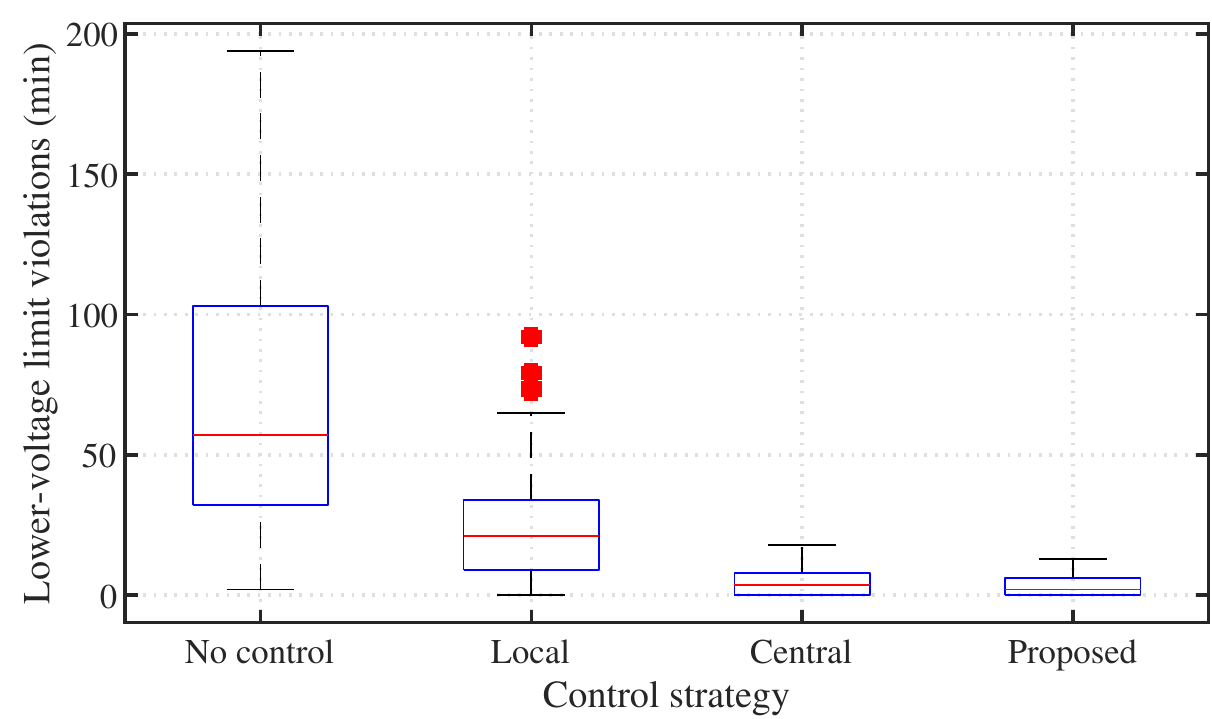}
\caption{Average daily lower-voltage limit violations under varying network settings with different control strategies.}
\label{Generalization}
\end{figure}

To test the extent to which the proposed approach can be generalized, we created 50 different network scenarios with varying inverter locations, quantities and capacities. In the construction of each case, 60 houses, 20--40 of which are equipped with smart PV inverter, are randomly sampled out of 167 candidate profiles from the Pecan Street data set. These houses are then randomly allocated to the 103-node network. For comparison, the local control strategy and the centralized organization scheme are also tested. In the simulations, as the over-voltage problem always occurs on lateral 3 and the abundant inverter reactive power resources of the small farm located at node 76 can effectively solve this problem, the over-voltage limit violation comparison results are not presented here. Fig.~\ref{Generalization} shows the average daily lower voltage limit violations. All three control strategies are able to improve the voltage profiles. However, as the local control suffers from inverter saturation problems, the average daily lower voltage limit violation time under this control is as high as 27.5 minutes. On the other hand, this value reduces to 4.6 minutes and 3.6 minutes when the centralized organization scheme or the proposed strategy is implemented. This is because these two strategies are able to coordinate the inverters for voltage regulation, and mitigate the insufficiency in local control resources. Furthermore, the proposed strategy achieves slightly better performance than the centralized organization scheme as it is able to organize the inverters in a more flexible way.

\section{Implementation in Unbalanced Networks}
\begin{figure*}[htbp!]
     \centering
     \begin{subfigure}[b]{0.3\textwidth}
         \centering
         \includegraphics[width=\textwidth]{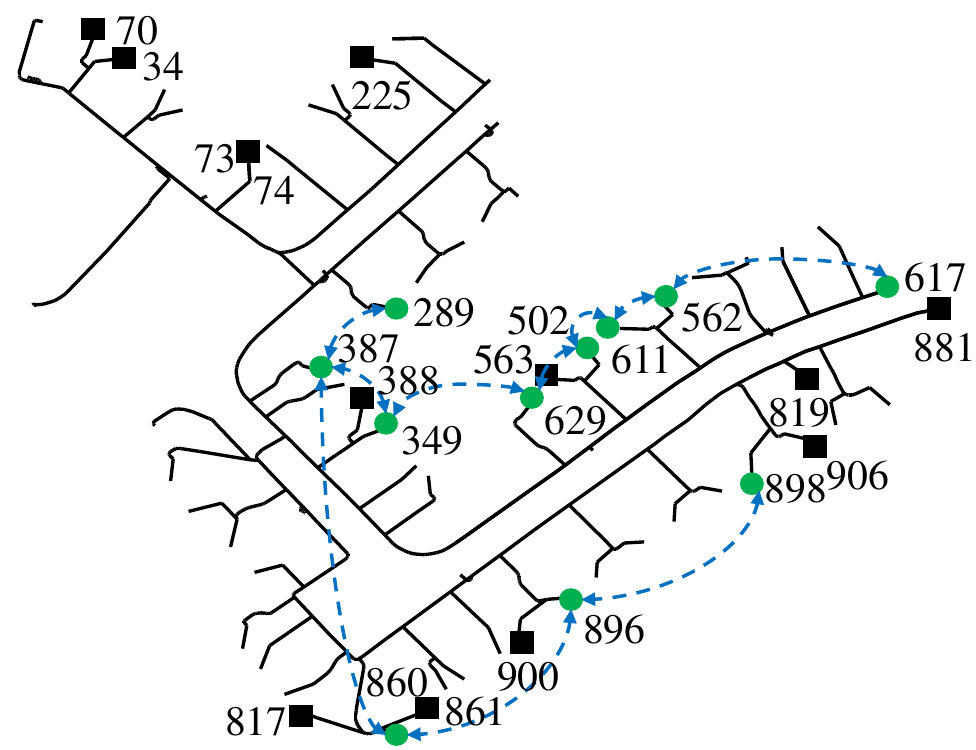}
         \caption{}
         \label{}
     \end{subfigure}
     \hfill
     \begin{subfigure}[b]{0.3\textwidth}
         \centering
         \includegraphics[width=\textwidth]{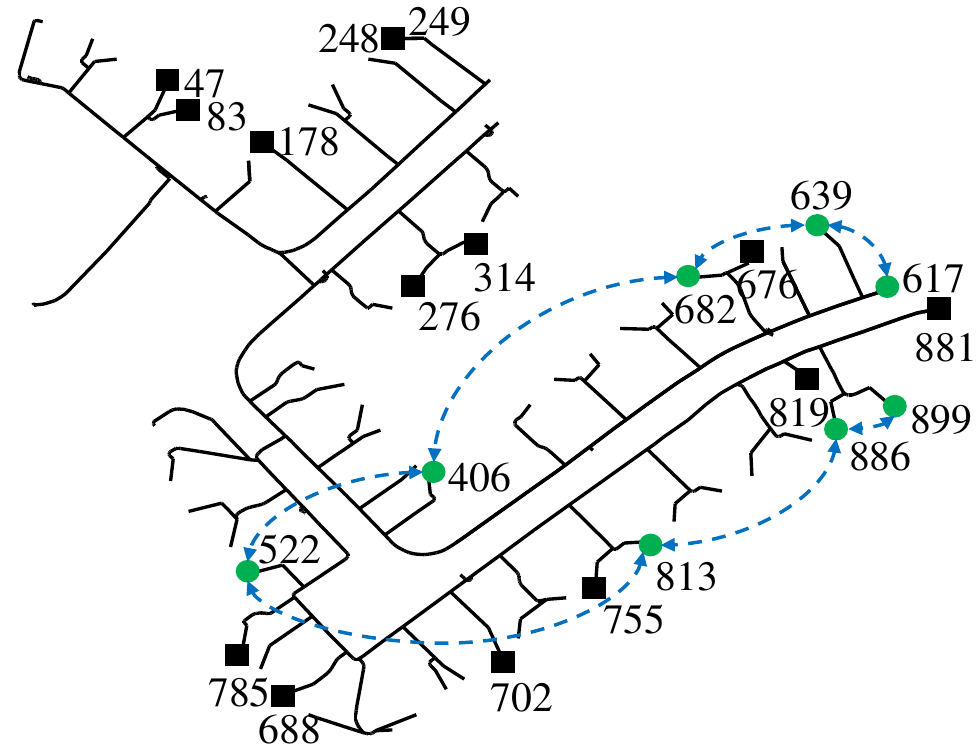}
         \caption{}
         \label{}
     \end{subfigure}
     \hfill
     \begin{subfigure}[b]{0.3\textwidth}
         \centering
         \includegraphics[width=\textwidth]{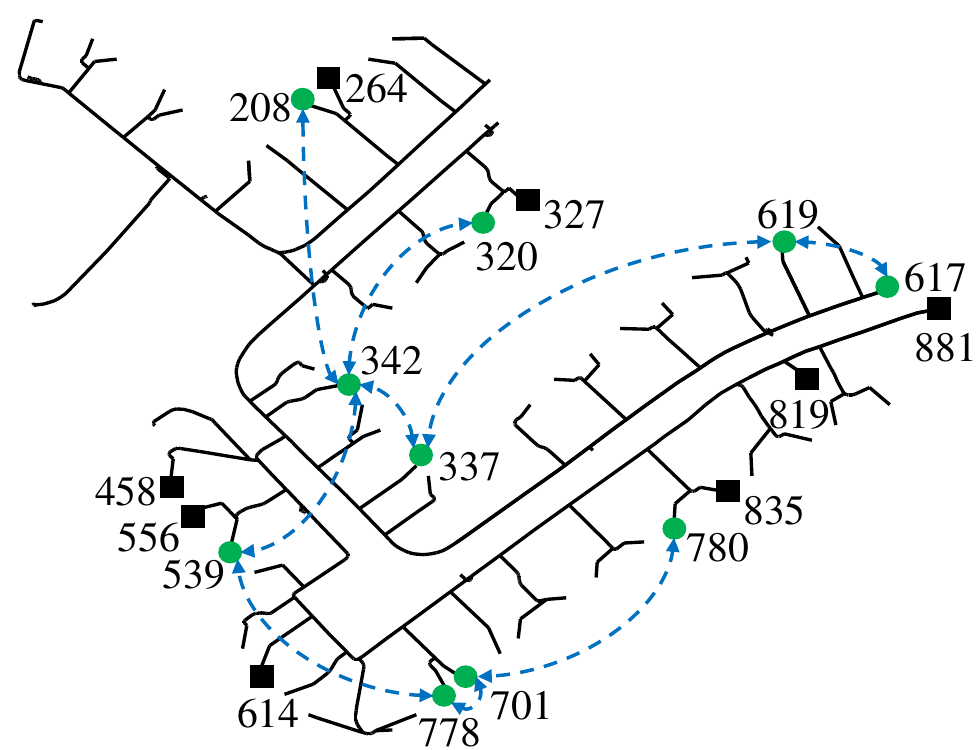}
         \caption{}
         \label{}
     \end{subfigure}
        \caption{Diagrams of the unbalanced test feeder. (a) Phase $a$, (b) Phase $b$, (c) Phase $c$}.
        \label{fig:Unbalance}
\end{figure*}

To assess the effectiveness of the proposed strategy in unbalanced networks, we modeled the IEEE European LV test feeder in OpenDSS \cite{OpenDSS}. The detailed network parameters and customer loading profiles can be found in \cite{LVN}. We extended
this network by connecting the bank and the office load described in Section VI-A at nodes 819 and 881. A three-phase small PV farm with a capacity of 28 kW at each phase is connected at node 617. The smart PV inverters are allocated to 10, 7 and 9 randomly selected houses on phases $a$, $b$ and $c$, respectively. The daily PV generation profiles are the same as described in Section VI-A. Fig.~\ref{fig:Unbalance} shows the modified test feeder. To maintain the voltage deviation within $\pm 5\%$ \cite{Consensus3, Consensus4}, we set the regulation range $[\bar{V}, \ubar{V}]$ as $[1.049, 0.951]$, the threshold range $[\bar{V}_\mathrm{th}, \ubar{V}_\mathrm{th}]$ as $[1.025, 0.975]$. The other parameters are as in Table~\ref{tab: control parameters}. The proposed control strategy is implemented in each phase separately.

\begin{figure}[htbp!]
\centering

\begin{subfigure}[b]{0.5\textwidth}
   \includegraphics[width=3.5in]{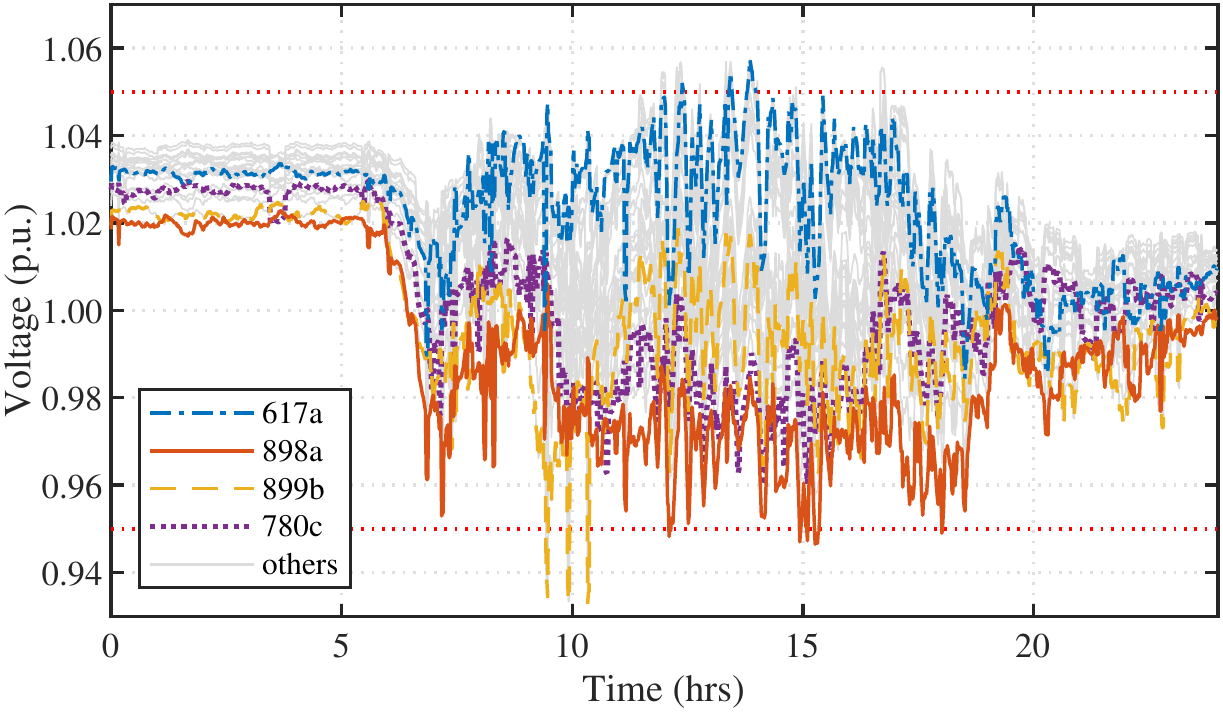}
   \caption{}
   \label{} 
\end{subfigure}

\begin{subfigure}[b]{0.5\textwidth}
   \includegraphics[width=3.5in]{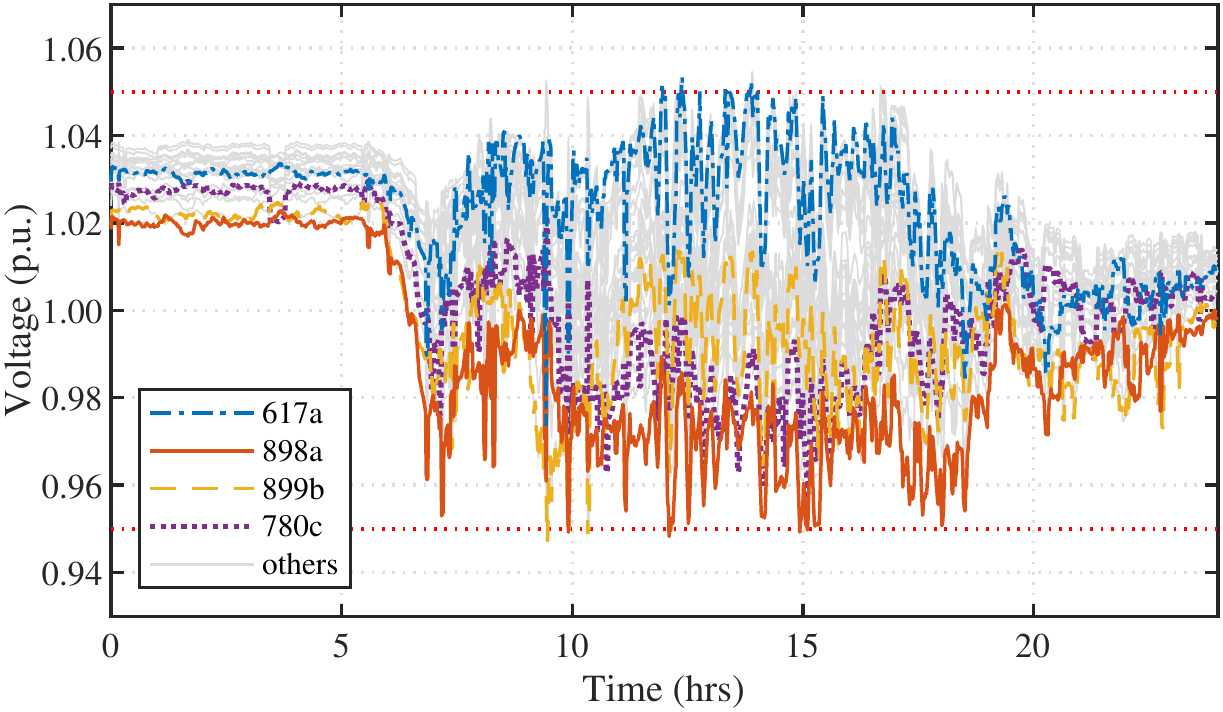}
   \caption{}
   \label{} 
\end{subfigure}

\caption{Voltage profiles of the network over a 24-hour period. (a) Without PV reactive power compensation; (b) With the proposed strategy.}
\label{ConVol3}
\end{figure}

\begin{figure}[htbp!]
\centering

\begin{subfigure}[b]{0.5\textwidth}
   \includegraphics[width=3.5in]{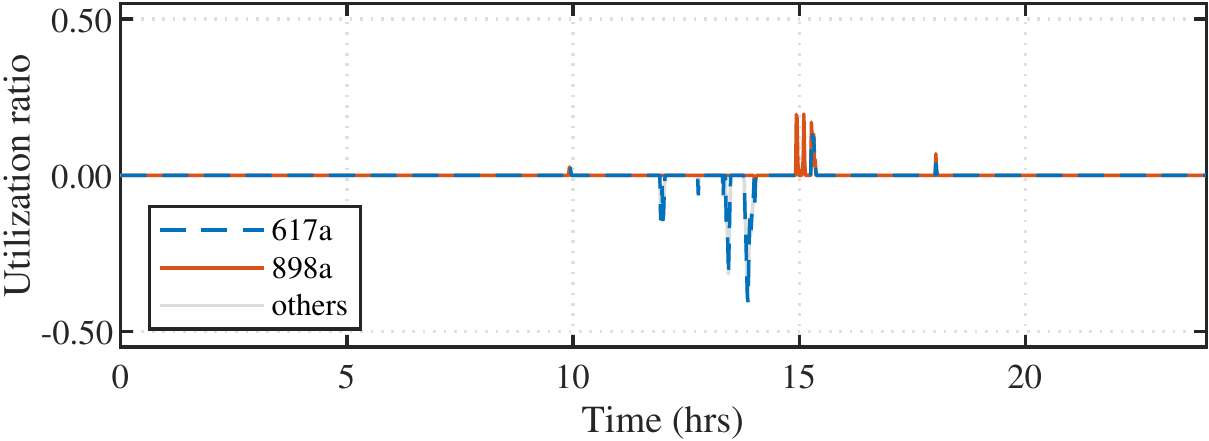}
   \caption{}
   \label{} 
\end{subfigure}

\begin{subfigure}[b]{0.5\textwidth}
   \includegraphics[width=3.5in]{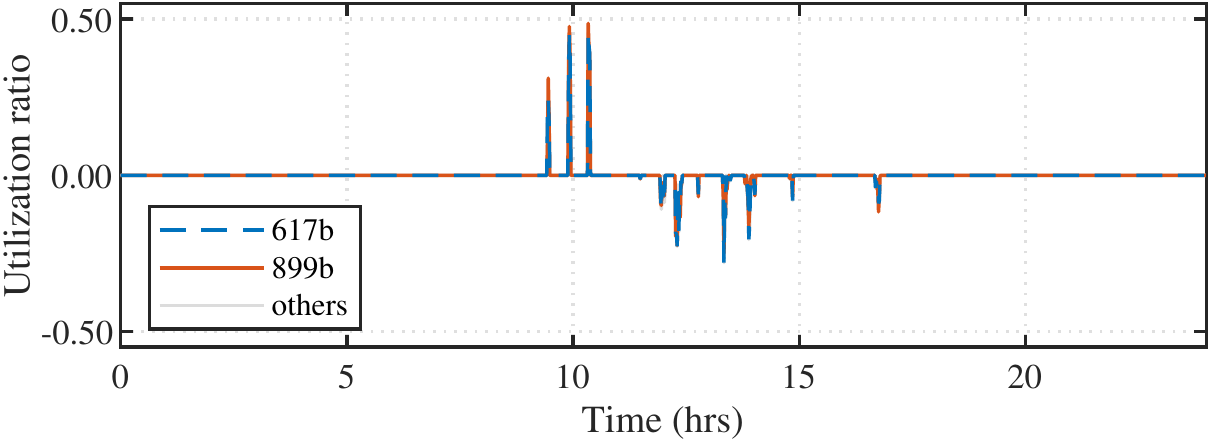}
   \caption{}
   \label{} 
\end{subfigure}

\begin{subfigure}[b]{0.5\textwidth}
   \includegraphics[width=3.5in]{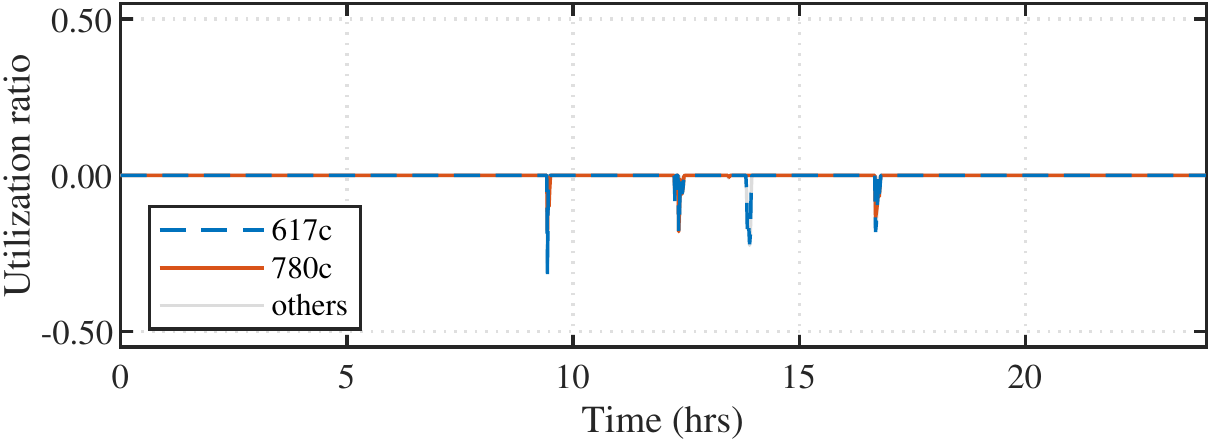}
   \caption{}
   \label{} 
\end{subfigure}

\caption{Utilization ratios of the smart PV inverter over a 24-hour period. (a) Phase $a$; (b) Phase $b$; (c) Phase $c$.}
\label{ConUR3}
\end{figure}

Fig.~\ref{ConVol3} demonstrates that the proposed strategy effectively improves the voltage profiles. Fig.~\ref{ConUR3} shows the coordinated actions of the smart PV inverters over the day. During the control process, the PV inverters at phase $a$ mainly form two coalitions since inverter 349 separate from inverter 629 at 10:45. These two coalitions are led by inverters 617 and 898 to solve their over-voltage and under-voltage problems. On the other hand, the PV inverters in phase $b$ remain in the same coalition the whole day while the coalition leader changes. For example, inverter 899 leads the coalition to eliminate the lower voltage limit violations from 9:25 to 10:25, while inverter 617 takes the leader position in the middle of the day to solve the over-voltage problems. The PV inverters on phase $c$ belong to the same coalition for most of the time. However, inverter 539 separates from inverter 342 at 13:45 and thus inverter 539, 778, 701 and 780 stop absorbing reactive power until 14:00. This happens as the voltage of these inverters are relatively low during this period.

\section{Conclusion and Future Works}
This paper developed and demonstrated a dynamic distributed control strategy to form coalition of PV inverters and their coordinated control for voltage regulation in distribution networks. This strategy is required when uneven integration of PV capacity creates large voltage differences along a feeder. On the slower time-scale, the coalition formation scheme separates opposing voltage violation problems and assigns them to groups of inverters with similar voltage margins and sufficient regulation capacity. On the faster time-scale, the feedback-based leader-follower consensus algorithm guides PV inverters within each coalition to eliminate the voltage violations while reaching a consensus on their utilization ratios. This control strategy effectively maintains voltages within the acceptable range while ensuring a sufficient and fair utilization of the available voltage regulation resources. 

One potential extension of the proposed approach is to incorporate the volt-watt control capability of the smart inverters. This extension would be particularly useful during the peak PV generation period, when the available reactive capacity of smart inverters is significantly reduced and the line congestion caused by reverse power flow becomes a more serious issue. Under these conditions, the extended approach would be able to promote an equal active power curtailment among customers, which would help mitigate line congestion. However, to achieve a more effective line congestion management, additional information about the network configuration and its operating status would be required \cite{Active}.

The proposed approach provides effective regulation when implemented on each phase separately in the unbalanced network case study. However, its full extension to the three-phase unbalanced networks merits further exploration. For example, three-phase and phase-to-phase connected PV inverters could be used to mitigate voltage imbalances, while the phase-to-neutral connected PV inverters regulate the voltage profiles \cite{Consensus4}. An extra inter-phase coordination signal could be communicated for the PV inverters to determine their reactive outputs \cite{Inter}. With these extensions, the coordination between inverters of different nodes, phases and connection types could be comprehensively considered.

Under the proposed approach, the smart PV inverters frequently exchange state information with one another, such as voltage magnitude and utilization ratio. Although the information exchange range is limited to direct neighbors, this process might still cause data privacy issues. To handle such problems, data encryption technology and advanced privacy-preserving consensus algorithms could be deployed \cite{Privacy1, Privacy2}. On the other hand, the smart inverters cooperate with each other based on the assumption that everyone is trustworthy and reliable, which may make them vulnerable to potential cyber-attacks such as malicious users and denial-of-service \cite{Security}. Further study is needed to improve the safety and reliability of the proposed approach.

Other interesting future research topics include accommodating different types of voltage regulation resources such as electric vehicles; designing the communication network structure for the coalition formation in meshed networks; and studying the convergence of the feedback-based leader-follower consensus algorithm under time-varying inverter capacities.

\ifCLASSOPTIONcaptionsoff
  \newpage
\fi


\begin{thebibliography}{99}
\bibitem{SolarPowerEurope}
Solar Power Europe, ``Global market outlook for solar power 2020-2024,'' Jun. 2020, [Online]. Available: \url{https://www.solarpowereurope.org/global-market-outlook-2020-2024/}
\bibitem{Voltagerise}
R.~Tonkoski, D.~Turcotte, and T.~H.~EL-Fouly, ``Impact of high PV penetration on voltage profiles in residential neighborhoods,'' \textit{IEEE Trans. Sustain. Energy}, vol. 3, no. 3, pp. 518-527, Jul. 2012.
\bibitem{Voltagestability}
R.~Yan, and T.~K.~Saha, ``Investigation of voltage stability for residential customers due to high photovoltaic penetrations,'' \textit{IEEE Trans. Power Syst.}, vol. 27, no. 2, pp. 651-662, May. 2012.
\bibitem{Voltagedifference}
L.~Wang, T.~K.~Saha and R.~Yan, ``Voltage regulation for distribution systems with uneven PV integration in different feeders,'' in \textit{Power and Energy Society General Meeting}, 2017 \textit{IEEE}, pp. 1-5.
\bibitem{Traditional1}
R.~Yan, B.~Marais, and T.~K.~Saha, ``Impacts of residential photovoltaic power fluctuation on on-load tap changer operation and a solution using DSTATCOM,'' \textit{Elect. Power Syst. Res.}, vol. 111, pp. 185-193, Jun. 2014.
\bibitem{Traditional2}
M.~I.~Hossain, R.~Yan, and T.~K.~Saha, ``Investigation of the interaction between step voltage regulators and large-scale photovoltaic systems regarding voltage regulation and unbalance,'' \textit{IET Renew. Power Gener.}, vol. 10, no. 3, pp. 299-309, Mar. 2016.
\bibitem{1547}
IEEE Std. 1547-2018: ``IEEE standard for interconnection and interoperability of distributed energy resources with associated electric power systems interfaces'', 2018
\bibitem{Local}
P.~Jahangiri, and D.~C.~Aliprantis, ``Distributed Volt/VAr control by PV inverters,'' \textit{IEEE Trans. Power Syst.}, vol. 28, no. 3, pp. 3429-3439, Aug. 2013.
\bibitem{Centralized}
A.~Kulmala, S.~Repo and P.~Järventausta: ``Coordinated voltage control in distribution networks including several distributed energy resources,'' \textit{IEEE Trans. Smart Grid}, vol. 5, no. 4, pp. 2010-2020, Jul. 2014.
\bibitem{Dopt1}
S.~Magnússon, G.~Qu and N.~Li, ``Distributed optimal voltage control with asynchronous and delayed communication,'' \textit{IEEE Trans. Smart Grid}, vol. 11, no. 4, pp. 3469-3482, July 2020.
\bibitem{Dopt2}
G.~Qu and N.~Li, ``Optimal distributed feedback voltage control under limited reactive power,'' \textit{IEEE Trans. Power Syst.}, vol. 35, no. 1, pp. 315-331, July 2019.
\bibitem{Consensus1}
G.~Mokhtari, A.~Ghosh, G.~Nourbakhsh and G.~Ledwich, ``Smart robust resources control in LV network to deal with voltage rise issue,'' \textit{IEEE Trans. Sustain. Energy}, vol. 4, no. 4, pp. 1043-1050, Oct. 2013.
\bibitem{Consensus2}
Y.~Wang, K.~T.~Tan, X.~Y.~Peng and P.~L.~So, ``Coordinated control of distributed energy-storage systems for voltage regulation in distribution networks,'' \textit{IEEE Trans. Power Del.}, vol. 31, no. 3, pp. 1132-1141, Jun. 2016.
\bibitem{Consensus3}
M.~Zeraati, M.~E.~H.~Golshan, and J.~M.~Guerrero, ``A consensus-based cooperative control of PEV battery and PV active power curtailment for voltage regulation in distribution networks,'' \textit{IEEE Trans. Smart Grid}, vol. 10, no. 1, pp. 670-680, Jan. 2019.
\bibitem{Consensus4}
M.~Zeraati, M.~E.~H.~Golshan, and J.~M.~Guerrero, ``Voltage quality improvement in low voltage distribution networks using reactive power capability of single-phase PV inverters,'' \textit{IEEE Trans. Smart Grid}, vol. 10, no. 5, pp. 5057-5065, Sep. 2019.
\bibitem{Consensus5}
G.~Mokhtari, G.~Nourbakhsh and A.~Ghosh, ``Smart coordination of energy storage units (ESUs) for voltage and loading management in distribution networks,'' \textit{IEEE Trans. Power Syst.}, vol. 28, no. 4, pp. 4812-4820, Nov. 2013.
\bibitem{Consensus6}
Y.~Wang, M.~H.~Syed, E.~Guillo-Sansano, Y.~Xu and G.~M.~Burt, ``Inverter-based voltage control of distribution networks: a three-level coordinated method and power hardware-in-the-loop validation,''\textit{IEEE Trans. Sustain. Energy}, vol. 11, no. 4, pp. 2380-2391, Oct. 2020.
\bibitem{Paradigm}
B.~Horling, and V.~Lesser, ``A survey of multi-agent organizational paradigms,'' \textit{Knowl. Eng. Rev.}, vol. 19, no. 4, pp. 281-316, Dec. 2004.
\bibitem{Smart1}
J.~W.~Smith, W.~Sunderman, R.~Dugan and B.~Seal, ``Smart inverter volt/var control functions for high penetration of PV on distribution systems,'' in \textit{Power Systems Conference and Exposition (PSCE)}, 2011 \textit{IEEE/PES}, pp. 1-6.
\bibitem{Smart2}
Y.~Xu, and J.~.M.~Guerrero, ``Smart inverter for utility and industry applications,'' in \textit{Proc. PCIM Eur. Int. Exhib. Conf. Power Electron., Intell. Motion, Renew. Energy Manage}, May. 2015, pp. 1-8.
\bibitem{PLC}
V.~C.~Gungor, and F.~C.~Lambert, ``A survey on communication networks for electric automation,'' \textit{Comput. Netw.}, vol. 50, no. 7, pp. 877-897, May. 2006.
\bibitem{maxconsensus}
R.~O.~Saber, and R.~M.~Murray, ``Consensus problems in networks of agents with switching topology and time-delays,'' \textit{IEEE. Trans. Autom. Control}, vol. 49, no. 9, pp. 1520-1533, Sept. 2004.
\bibitem{Schedule}
D.~Ye, M.~Zhang, and D.~Soetanto, ``Decentralized dispatch of distributed energy resources in smart grids via multi-agent coalition formation,'' \textit{J. Parallel Distrib. Comput.}, vol. 83, pp. 30-43, 2015.
\bibitem{Trading}
F.~Luo, Z.~Y.~Dong, G.~Liang, J.~Murata and Z.~Xu, ``A distributed electricity trading system in active distribution networks based on multi-agent coalition and blockchain,'' \textit{IEEE Trans. Power Syst.}, vol. 34, no. 5, pp. 4097-4108, Sept. 2019.
\bibitem{Restoration}
F.~Ren, M.~Zhang, D.~Soetanto and X.~Su, ``Conceptual design of a multi-agent system for interconnected power systems restoration,'' \textit{IEEE Trans. Power Syst.}, vol. 27, no. 2, pp. 732-740, May 2012.
\bibitem{Partition1}
B.A.~Faiya, D.~Athanasiadis, M.~J.~Chen, S.~McArthur, I.~Kockar, H.~Lu and F.~de Le\'on, ``A self organizing multi agent system for distributed voltage regulation,'' \textit{IEEE Trans. Smart Grid}, 2021.
\bibitem{Partition2}
P.~Li, C.~Zhang, Z.~Wu, Y.~Xu, M.~Hu and Z.~Dong, ``Distributed adaptive robust voltage/VAR control with network partition in active distribution networks,'' \textit{IEEE Trans. Smart Grid}, vol. 11, no. 3, pp. 2245-2256, May 2020.
\bibitem{Consensus}
R.~Olfati-Saber, J.~A.~Fax, and R.~M.~Murray, ``Consensus and cooperation in networked multi-agent systems,'' \textit{Proc. IEEE}, vol. 95, no. 1, pp. 215-233, Jan. 2007.
\bibitem{Convergence1}
C.~Chang, M.~Colombino, J.~Corté and E.~Dall’Anese, ``Saddle-flow dynamics for distributed feedback-based optimization,'' \textit{IEEE Control Syst. Lett.}, vol. 3, no. 4, pp. 948-953, Oct. 2019.
\bibitem{Cooperative}
Toronto and Region Conservation Authority (TRCA) and York University, ``Managing Volt/VAR in active distribution networks using distributed generation units and peer to peer communication,'' Toronto and Region Conservation Authority, Vaughan, Ontario, 2018.
\bibitem{Incentive}
C.~Wu, G.~Hug and S.~Kar, ``Smart inverter for voltage regulation: physical and market implementation,'' \textit{IEEE Trans. Power Syst.}, vol. 33, no. 6, pp. 6181-6192, Nov. 2018.
\bibitem{Ownership}
M.~Joyce, ``Advanced inverter functions to support high levels of distributed solar,'' National Renewable Energy Laboratory, 2015.
\bibitem{Network}
B.~Wei, Z.~Qiu and G.~Deconinck, ``A mean-field voltage control approach for active distribution networks with uncertainties,'' \textit{IEEE Trans. Smart Grid}, vol. 12, no. 2, pp. 1455-1466, Mar. 2021.
\bibitem{Pecan}
(2019). \textit{Pecan street Inc. Dataport}. [Online]. Available: \url{https://dataport.pecanstreet.org/}
\bibitem{EnerNOC}
EnerNOC Open Data. [Online]. Available: \url{https://open-enernoc-data.s3.amazonaws.com/anon/index.html}
\bibitem{MATPOWER}
R.~D.~Zimmerman, C.~E.~Murillo-Sánchez and R.~J.~Thomas, ``MATPOWER: steady-state operations, planning, and analysis tools for power systems research and education," \textit{IEEE. Trans. Power Syst.}, vol. 26, no. 1, pp. 12-19, Feb. 2011.
\bibitem{SOCP}
M.~Farivar, and S.~H.~Low, ``Branch flow model: relaxations and convexification,'' \textit{IEEE. Trans. Power Syst.}, vol. 28, no. 3, pp. 2554-2564, Aug. 2013.
\bibitem{Integral}
N.~Li, G.~Qu, and M.~Dahleh, ``Real-time decentralized voltage control in distribution networks,'' in \textit{Proc. 52nd Annu. Allerton Conf. Commun. Control Comput.}, Monticello, IL, USA, 2014, pp.582-588.
\bibitem{Lifetime}
O.~Gandhi, C.~D.~Rodríguez-Gallegos, N.~B.~Y.~Gorla, M.~Bieri, T.~Reindl and D.~Srinivasan, ``Reactive power cost from PV inverters considering inverter lifetime assessment,'' \textit{IEEE Trans. Sustain. Energy}, vol. 10, no. 2, pp. 738-747, Apr. 2019.
\bibitem{OpenDSS}
R.~C.~Dugan, ``Reference guide: the open distribution system simulator (openDSS),'', Elect. Power Res. Inst., Palo Alto, CA, USA, 2012.
\bibitem{LVN}
IEEE PES Distribution Systems Analysis Subcommittee Radial Test Feeders. [Online]. Available: \url{https://site.ieee.org/pes-testfeeders/resources/}
\bibitem{Active}
A.~T.~Procopiou and L.~F.~Ochoa, ``Asset congestion and voltage management in large-scale MV-LV networks with solar PV,'' \textit{IEEE Trans. Power Syst.}, vol. 36, no. 5, pp. 4018-4027, Sept. 2021.
\bibitem{Inter}
L.~Wang, R.~Yan, F.~Bai, T.~Saha and K.~Wang, ``A distributed inter-phase coordination algorithm for voltage control with unbalanced PV integration in LV systems,'' \textit{IEEE Trans. Sustain. Energy}, vol. 11, no. 4, pp. 2687-2697, Oct. 2020.
\bibitem{Privacy1}
H.~Xu, Y.~-H.~Ni, Z.~Liu and Z.~Chen, ``Privacy-preserving leader-following consensus via node-augment mechanism,'' \textit{IEEE Trans. Circuits Syts. II, Exp. Briefs}, vol. 68, no. 6, pp. 2117-2121, June 2021.
\bibitem{Privacy2}
X.~Duan, J.~He, P.~Cheng, Y.~Mo, and J.~Chen, ``Privacy preserving maximum consensus,'' in \textit{Proc. IEEE 54th Annu. Conf. Decis. Control (CDC)}, 2015, pp. 4517–4522.
\bibitem{Security}
P.~Li, Y.~Liu, H.~Xin, and X.~Jiang, “A robust distributed economic dispatch strategy of virtual power plant under cyber-attacks,” \textit{IEEE Trans. Ind. Inform.}, vol. 14, no. 10, pp. 4343–4352, Oct. 2018.
\end{thebibliography}
\end{document}